\documentclass[a4paper,11pt]{article}

\usepackage{draft}
\usepackage{orcidlink}
\usepackage{mathtools}
\usepackage{cite}
\usepackage{bm}
\definecolor{twilightlavender}{rgb}{0.54, 0.29, 0.42}
\definecolor{richmaroon}{rgb}{0.69, 0.19, 0.38}
\definecolor{forestgreen(web)}{rgb}{0.13, 0.55, 0.13}
\definecolor{lava}{rgb}{0.81, 0.06, 0.13}
\hypersetup{
	breaklinks,
	colorlinks,
	citecolor=forestgreen(web),
	filecolor=richmaroon,
	linkcolor=lava,
	urlcolor=twilightlavender
}

\title{Path integral quantization of null bosonic strings with Carroll-Weyl ghosts}

\affiliation[a]{Yau Mathematical Sciences Center (YMSC), Tsinghua University, Beijing 100084, China}
\affiliation[b]{Harish-Chandra Research Institute, A CI of Homi Bhabha National Institute,
Chhatnag Road, Jhunsi, Prayagraj (Allahabad), Uttar Pradesh 211019, India}

\author[a,\orcidlink{0000-0002-4535-3198}]{Sarthak Duary}\emailAdd{sarthakduary@tsinghua.edu.cn}
\author[b,\orcidlink{0009-0008-2963-2497}]{and Sourav Maji}\emailAdd{souravmaji@hri.res.in}

\abstract{We revisit the path integral quantization of the null bosonic string from the viewpoint that all local gauge symmetries of the Carrollian worldsheet must be gauge fixed before the quantum theory is defined. In the tensile-string construction the $bc$ ghosts are the Faddeev-Popov determinant for fixing $\mathrm{Diff}\times\mathrm{Weyl}$. In the ILST null string this logic gives the BMS $bc$ system. However, a Carrollian worldsheet admits an additional volume-preserving Carroll-Weyl scaling, whose Hamiltonian generator is $C_3=P\cdot X$. Keeping this scaling as a genuine local gauge symmetry adds one more Faddeev-Popov row. The correct ghost system is therefore a $bcs$ system: the BMS $bc$ ghosts plus a scalar ghost $s$ and scalar antighost $b^s$ for Carroll-Weyl scaling. We derive the revised path integral, the $bcs$-ghost action, its residual symmetry equations, mode expansion, and its relation to the extended BMS algebra. The result changes the BRST complex and the anomaly problem: the usual $D=26$ check based only on the old BMS $bc$ ghosts is a partially gauge-fixed calculation, while the Carroll-Weyl covariant quantum theory must include the $s,b^s$ sector.}

\newcommand{\dd}{\mathrm d}

\newcommand{\cD}{\mathcal D}
\newcommand{\cW}{\mathcal W}
\newcommand{\cM}{\mathcal M}

\newcommand{\gh}{\mathrm{gh}}
\newcommand{\CW}{\mathrm{CW}}
\newcommand{\diff}{\mathrm{diff}}
\newcommand{\Diff}{\mathrm{Diff}}
\newcommand{\Weyl}{\mathrm{Weyl}}
\newcommand{\Vol}{\mathrm{Vol}}
\newcommand{\FP}{\mathrm{FP}}

\begin{document}
\maketitle

\section{Introduction}
The guiding principle of the Polyakov path integral is that every local gauge redundancy must be removed from the functional integral, and the associated Faddeev-Popov Jacobian constitutes an intrinsic part of the quantum theory.
In the tensile-string construction the starting point is the formal integral over embeddings and metrics,
\([dX\,dg]/\Vol(\Diff\times\Weyl)\), and gauge fixing the metric to a fiducial representative
\(\hat g_{ab}\) produces the Faddeev-Popov determinant \(\Delta_{\FP}(\hat g)\).  The determinant is
not an auxiliary decoration; it is the \(bc\) conformal field theory whose central charge cancels the
matter central charge only in the critical dimension.  This is the path-integral origin of the familiar
statement \(c_X+c_{\gh}=D-26\).

The tensionless limit is not a small technical deformation of the tensile theory.  It is the
\(\alpha'\to\infty\), or \(T\to0\), regime in which the high-energy string amplitudes studied by
Gross and Mende simplify and exhibit large hidden symmetry structures
\cite{Gross:1987kza,Gross:1987ar,Gross:1988ue}.  In this limit the infinite tower of massive string
states becomes massless, making tensionless strings a natural laboratory for higher-spin symmetry and
holography \cite{Sundborg:2000wp,Vasiliev:2003ev}.  In flat space the same limit gives the null strings
introduced by Schild \cite{Schild:1976vq}, later developed in the ILST formalism
\cite{Isberg:1993av,Isberg:1992ia}. The zero-tension
limit of the classical bosonic string was independently studied in \cite{Karlhede:1986wb},
where Dirac space was
introduced as a natural arena for this limit.  This approach was later
generalized to the conformal string in \cite{Gustafsson:1994kr}. The modern importance of the ILST description is that its temporal gauge
residual symmetry is BMS\(_3\), or equivalently the two-dimensional Galilean conformal algebra
\cite{Bagchi:2013bga, Bagchi:2015nca}, connecting null strings to BMS field theory, flat holography, and ambitwistor
strings \cite{Casali:2016atr,Casali:2017zkz,Mason:2013sva}.

The tensionless bosonic string should therefore be treated with the same path-integral discipline as
the Polyakov string.  In the ILST null-string action \cite{Isberg:1993av}, the Polyakov density
\(\sqrt{-g}\,g^{ab}\) is replaced by the rank-one Carrollian density \(V^aV^b\).  The path-integral analysis of
\cite{Chen:2023esw} fixes \(V^a=(1,0)\) and obtains a BMS version of the \(bc\) ghosts.  This is the
right analogue of the tensile-string construction if the only gauge redundancy left to fix is the
diffeomorphism redundancy of \(V^a\).  The physical constraints generated by this older ghost system
are the BMS pair \(P^2=0\) and \(P\cdot X'=0\).

The motivation for the present paper is the important recent observation of
Sheikh-Jabbari and Yavartanoo
\cite{SheikhJabbari:2026Consistency} that the Carrollian worldsheet of a null
string possesses a richer gauge structure than is manifest in the traditional
tensionless-limit treatment. Historically, the null string has been described by the two first-class
constraints
\(
C_1=P^2, C_2=P\cdot X',
\)
leading to the familiar counting of \((D-2)\) propagating target-space degrees
of freedom. Recent analyses
\cite{SheikhJabbari:2026Consistency,SheikhJabbari:2026I,SheikhJabbari:2026II}
have argued that the ordinary ILST null string possesses an additional
Carroll-Weyl gauge symmetry generated by
\(
C_3=P\!\cdot\!X,
\)
so that the complete constraint system is
\(
C_1=P^2, C_2=P \cdot X', C_3=P \cdot X.
\)
The extra gauge symmetry removes one further local degree of freedom,
modifying the classical counting from \((D-2)\) to \((D-3)\). This proposal is
distinct from the earlier conformal-string construction
 \cite{Gustafsson:1994kr,Lindstrom:2026quz} which is formulated in an enlarged
target space and includes additional constraints. 
 

The geometric origin of this symmetry was given in the gauged
Carroll-Weyl formulation of \cite{SheikhJabbari:2026I}.  A central point of
that work is that a Carrollian worldsheet admits two independent Weyl-type
scalings, unlike an ordinary Lorentzian Polyakov worldsheet, which has only
the usual Weyl rescaling.  One linear combination is the standard
volume-modulating Carroll-Weyl scaling, while the second is a
volume-preserving Carroll-Weyl scaling with no analogue in the non-degenerate
Lorentzian worldsheet geometry.  The latter is precisely the geometric origin
of the additional $\chi$ symmetry.  In the fully gauged description of
\cite{SheikhJabbari:2026I}, this symmetry is implemented by introducing a
Carroll-Weyl connection $W_a$ and replacing ordinary derivatives by
Carroll-Weyl covariant derivatives.  The gauged action is then invariant
under worldsheet diffeomorphisms together with the two Carroll-Weyl scalings.
Upon imposing the gauge condition $V^a W_a=0$, one recovers the ILST action,
but the residual part of the Carroll-Weyl symmetry survives precisely as the
codimension-one $\chi(\sigma)$ symmetry identified in
\cite{SheikhJabbari:2026Consistency}.

The necessity of this enlarged gauge structure was further established in the
consistent classical and Hamiltonian analysis of
\cite{SheikhJabbari:2026II}.  In that treatment the complete set of local
constraints is
\begin{equation}
  C_1=P^2,\qquad
  C_2=P\cdot X',\qquad
  C_3=P\cdot X .
\end{equation}
Here $C_1$ and $C_2$ are the usual null-string constraints, while
$C_3$ is the additional constraint generated by the Carroll-Weyl
$\chi$ symmetry.  Equivalently, the standard BMS$_3$ constraint algebra is
enlarged by a weight-one generator associated with $C_3$.  Thus, in the
Carroll-Weyl covariant formulation, omitting $C_3$ is not merely a different
choice of notation or gauge; it amounts to not quotienting by all local gauge
orbits of the null-string worldsheet theory.

We therefore distinguish carefully between the complete Carroll-Weyl
covariant formulation developed in
\cite{SheikhJabbari:2026Consistency,SheikhJabbari:2026I,SheikhJabbari:2026II}
and earlier analyses based on the traditional tensionless-limit formulation,
including the ILST framework and subsequent related discussions.  While the
earlier literature contains important ingredients of null-string theory, the
specific $\chi(\sigma)$ gauge symmetry and its associated constraint $C_3$
were first identified and emphasized in
\cite{SheikhJabbari:2026Consistency}, given a first-principles Carrollian
geometric origin in \cite{SheikhJabbari:2026I}, and treated systematically at
the classical Hamiltonian level in \cite{SheikhJabbari:2026II}.  Our analysis
is based on this gauge-complete viewpoint.  In particular, the ghost system
constructed below should be understood as the Faddeev-Popov system obtained
after gauge fixing the full local symmetry group, including the
Carroll-Weyl $\chi$ symmetry generated by $C_3$.

The new constraint \(C_3\) removes the local scale of the embedding in target space and extends the
BMS algebra by a weight-one current.  Omitting it in the path integral is therefore not a harmless
choice of notation: it means that one has not divided by all local gauge orbits.

Our central claim is the path-integral counterpart of this Hamiltonian statement.  Once
Carroll-Weyl scaling is kept as a local gauge symmetry, the Faddeev-Popov determinant has three rows,
not two.  The two-component BMS \(bc\)-ghost system is promoted to\footnote{We adopt the notation `\(bcs\)', following \cite{SheikhJabbari:2026II}.}
\begin{equation}
  (b,c)\quad\longrightarrow\quad (b,c,s),
\end{equation}
where \(s\) is the fermionic scalar ghost for Carroll-Weyl transformations.  Equivalently, the
antighost sector also acquires a scalar component, denoted \(b^s\), and the gauge-fixed theory contains
\begin{equation}
  b^A=(b^0,b^1,b^s),\qquad c^A=(c^0,c^1,s).
\end{equation}
We call this the BMS-Carroll \(bcs\)-ghost system.  Its most important term is not the new kinetic
term alone but the algebraic mixing \(s\,b^0\), which remembers that the temporal component
\(V^0\) itself transforms under Carroll-Weyl scaling.  This mixing is the path-integral reflection of
the non-trivial bracket between the Carroll-Weyl generator and the supertranslation constraint.

\paragraph{Main results.}
The first main result is that the Carroll-Weyl covariant Faddeev-Popov operator is the local
three-row operator
\begin{equation}
  \cM_{\CW}
  =
  \begin{pmatrix}
    -\frac12\partial_0 & \frac12\partial_1 & -1\\
    0 & -\partial_0 & 0\\
    0 & 0 & -\partial_0
  \end{pmatrix},
  \label{intro-main-M}
\end{equation}
acting on \((\epsilon^0,\epsilon^1,\lambda)\).  The old BMS \(bc\) determinant is only the upper-left
block of \eqref{intro-main-M}.  The second main result is the local ghost action
\begin{equation}
  S_{\mathrm{gf}}
  =
  \frac{1}{2\pi}\int\dd^2\sigma\,
  \left[
  \dot X^2
  +i\left(
  c^0\partial_0b^0
  -c^1\partial_1b^0
  +2c^1\partial_0b^1
  +s\partial_0b^s
  -2sb^0
  \right)
  \right],
  \label{intro-main-action}
\end{equation}
whose last two terms are forced by the Carroll-Weyl gauge fixing.  The third main result is the
modified ghost evolution
\begin{equation}
  \partial_0c^0-\partial_1c^1+2s=0,
  \qquad
  \partial_0b^s-2b^0=0,
  \label{intro-main-ghost-eom}
\end{equation}
which shows explicitly that the \(s,b^s\) sector is coupled to the BMS temporal ghosts.  Finally, the
Hamiltonian constraints close into the extended BMS algebra containing
\begin{equation}
  \{L_m,S_n\}=inS_{m+n},
  \qquad
  \{M_m,S_n\}=-2M_{m+n}.
  \label{intro-main-extended}
\end{equation}
The physics is therefore simple: the Carroll-Weyl ghost is the path-integral variable that removes the
local target-space scale generated by \(P\cdot X\).  It changes the BRST complex, the ghost zero-mode
measure, the construction of vertex operators, and the anomaly problem for null-string scattering.

This has immediate physical implications.  First, the BRST charge must contain the ghost for
\(C_3\), so the old BMS BRST operator is a partially gauge-fixed object.  Second, the anomaly
problem must be posed for the extended BMS algebra, not only for its \(L_n,M_n\) subalgebra.  Third,
vertex operators and physical-state conditions must be invariant under the additional local scaling.
These three statements are the reason the present paper is foundational: before one computes spectra,
vertex operators, amplitudes, or critical dimensions, the quantum measure must first know the complete
gauge orbit being divided out.  The Carroll-Weyl ghost is the missing measure degree of freedom.

\paragraph{Organization of the paper.}
The paper is organized as follows.  Section \ref{sec:prelim} reviews the usual ILST path integral and
the BMS \(bc\) ghosts, starting from the tensile-string \(P_1\) operator and then taking the null-string
analogue.  Section \ref{sec:cw} introduces the Carroll-Weyl gauged null-string action and its
constraints.  Section \ref{sec:fp-bcs} gives the revised Faddeev-Popov determinant and derives the
\(bcs\)-ghost action step by step.  Section \ref{sec:modes} derives the equations of motion and mode
expansions of the new ghost system.  Section \ref{sec:extended} discusses the extended BMS algebra,
BRST structure, and anomaly implications.  Section \ref{sec:conclusion} gives conclusions and future
directions. Appendix \ref{app:fp-details} collects the detailed Faddeev-Popov algebra. Appendix
\ref{app:ghost-action-details} derives the ghost action component by component. Appendix
\ref{app:modes-details} gives the derivation of the \(bcs\) equations and modes. Appendix \ref{app:oscillatoralgebra} presents detailed derivation of the ghost oscilator algebra. Appendix \ref{app:semidirect} explains the semi-direct product in the measure of the path integral due to the Carroll-Weyl gauge group. Appendix \ref{app:grassmann-det} gives the Grassmann proof of the fermionic determinant identity.

\section{Preliminaries: ILST path integral and BMS \texorpdfstring{\(bc\)}{bc} ghosts}
\label{sec:prelim}

\subsection{Faddeev-Popov determinant in tensile string theory}

We begin with the Faddeev-Popov determinant in tensile string theory construction, because the point of the present paper is a direct
application of the same rule to a Carrollian gauge symmetry.  The Euclidean
Polyakov path integral is \cite{TextbookStringTheory}
\begin{equation}
  Z[\hat g]
  =
  \int \frac{[dX\,dg]}{\Vol(\Diff\times\Weyl)}
  \exp[-S_X[X,g]],
  \qquad
  S_X
  =
  \frac{1}{4\pi\alpha'}\int_M\dd^2\sigma\,g^{1/2}
  g^{ab}\partial_aX^\mu\partial_bX_\mu .
  \label{polyakov-pol}
\end{equation}
The gauge slice is \(g_{ab}=\hat g_{ab}\), and the Faddeev-Popov identity is written as following
\begin{equation}
  1
  =
  \Delta_{\FP}(g)
  \int[d\zeta]\,\delta(g-\hat g^\zeta),
  \label{pol-fp}
\end{equation}
as in \cite{TextbookStringTheory}.
Here \(\zeta\) denotes a combined diffeomorphism and Weyl transformation.  Inserting
\eqref{pol-fp} and using gauge invariance leaves the measure on the slice,
\begin{equation}
  Z[\hat g]
  =
  \int[dX]\,\Delta_{\FP}(\hat g)\,
  \exp[-S_X[X,\hat g]] .
  \label{pol-slice}
\end{equation}

Near the identity, with \(\delta\sigma^a\) the vector generating diffeomorphisms and
\(\delta\omega\) the Weyl parameter, the metric variation decomposes as
\begin{equation}
  \delta g_{ab}
  =
  (2\delta\omega-\nabla_c\delta\sigma^c)g_{ab}
  -2(P_1\delta\sigma)_{ab},
  \label{metric-var-pol}
\end{equation}
where
\begin{equation}
  (P_1\delta\sigma)_{ab}
  =
  \frac12\left(
  \nabla_a\delta\sigma_b+\nabla_b\delta\sigma_a
  -g_{ab}\nabla_c\delta\sigma^c
  \right).
  \label{P1-pol}
\end{equation}
The Weyl part fixes the trace of \(\delta g_{ab}\), while \(P_1\) maps a vector into a traceless
symmetric tensor.  This is why the ordinary tensile antighost \(b_{ab}\) is traceless.  Replacing the
bosonic variables in the inverse determinant by Grassmann variables gives
\begin{equation}
  \Delta_{\FP}(\hat g)
  =
  \int[db\,dc]\exp[-S_g],
  \qquad
  S_g
  =
  \frac{1}{2\pi}\int\dd^2\sigma\,\hat g^{1/2}
  b_{ab}(\hat P_1c)^{ab}.
  \label{pol-ghost-action}
\end{equation}
This is the ghost determinant.
Equivalently, using tracelessness,
\begin{equation}
  S_g
  =
  \frac{1}{2\pi}\int\dd^2\sigma\,\hat g^{1/2}
  b_{ab}\hat\nabla^ac^b .
  \label{pol-ghost-action2}
\end{equation}
In conformal coordinates this becomes
\begin{equation}
  S_g
  =
  \frac{1}{2\pi}\int\dd^2z\,
  \left(
  b_{zz}\partial_{\bar z}c^z
  +b_{\bar z\bar z}\partial_zc^{\bar z}
  \right),
  \label{pol-chiral-bc}
\end{equation}
so \(b\) has weight \(2\), \(c\) has weight \(-1\), and \(c_X+c_{\gh}=D-26\), as in \cite{TextbookStringTheory}.

There are two lessons we will use below.  First, the ghost system is determined by the complete set of
gauge conditions and gauge parameters, not by analogy with a known CFT.  Second, if the gauge algebra
is enlarged, the Faddeev-Popov operator is enlarged and the ghost complex changes.

\subsection{The BMS \texorpdfstring{\(bc\)}{bc} ghosts of the ILST path integral}

We now recall the part of the tensionless-string analysis that will survive as a subsector of the
Carroll-Weyl covariant construction.  Worldsheet indices are denoted
\(a,b,c,\ldots=0,1\), with \(\sigma^0=\tau\), \(\sigma^1=\sigma\), and
\(\sigma\sim\sigma+2\pi\).  The ILST action is
\begin{equation}
  S_{\mathrm{ILST}}[V,X]
  =
  \frac{1}{2\pi}\int \dd^2\sigma\,
  V^aV^b\partial_aX\cdot\partial_bX .
  \label{ilst}
\end{equation}
It is the Carrollian replacement
of the Polyakov kinetic density \(g^{1/2}g^{ab}\partial_aX\partial_bX\): the non-degenerate
inverse metric density is replaced by the rank-one density \(V^aV^b\).
The Carrollian inverse metric density is rank one,
\begin{equation}
  \sqrt{-g}\,g^{ab}=V^aV^b .
\end{equation}
In temporal gauge,
\begin{equation}
  V^a=(1,0),
  \label{temporalV}
\end{equation}
the matter action becomes
\begin{equation}
  S_{\mathrm{ILST}}\big|_{V^{a}=(1,0)}
  =
  \frac{1}{2\pi}\int\dd^2\sigma\,\dot X^2 .
\end{equation}

The vector density \(V^a\) has weight \(+1/2\) in the convention used here.  Under an infinitesimal
diffeomorphism generated by \(\epsilon^a\),
\begin{equation}
  \delta_\epsilon V^a
  =
  -V^b\partial_b\epsilon^a
  +\epsilon^b\partial_bV^a
  +\frac12V^a\partial_b\epsilon^b .
  \label{deltaV-diff}
\end{equation}
Putting \(V^a=(1,0)\), one finds
\begin{align}
  \delta_\epsilon V^0
  &=
  -\frac12\partial_0\epsilon^0+\frac12\partial_1\epsilon^1,
  \\
  \delta_\epsilon V^1
  &=
  -\partial_0\epsilon^1 .
\end{align}
The residual transformations obey
\begin{equation}
  \partial_0\epsilon^1=0,\qquad
  \partial_0\epsilon^0=\partial_1\epsilon^1,
\end{equation}
so
\begin{equation}
  \epsilon^0=f'(\sigma)\tau+g(\sigma),\qquad
  \epsilon^1=f(\sigma),
\end{equation}
which is the BMS form.

The ILST path integral before gauge fixing is formally
\begin{equation}
  Z_{\mathrm{ILST}}
  =
  \int \frac{DV\,DX}{\mathrm{Vol}(\diff)}\,\exp\{iS_{\mathrm{ILST}}[V,X]\}.
\end{equation}
Inserting the Faddeev-Popov identity for the gauge \(V^a=(1,0)\),
\begin{equation}
  1
  =
  \int D\epsilon\,
  \delta(V^\epsilon-V_{\mathrm{gauge}})
  \det\!\left(\frac{\delta V^b}{\delta\epsilon^a}\right),
\end{equation}
leads to the operator
\begin{equation}
  \Delta^b{}_a
  =
  \frac{\delta V^b}{\delta\epsilon^a}
  =
  -\delta^b{}_aV^c\partial_c+\partial_aV^b+\frac12V^b\partial_a .
  \label{Delta-old}
\end{equation}
This is the BMS analogue of the \(P_1\) operator in \eqref{P1-pol}.  It maps infinitesimal
diffeomorphisms into deformations of the gauge-fixed Carrollian density \(V^a\), and it is the
operator denoted \(\Delta^b{}_a\) in \cite{Chen:2023esw}.
On \(V^a=(1,0)\),
\begin{equation}
  \Delta^b{}_a
  =
  \begin{pmatrix}
    -\frac12\partial_0 & \frac12\partial_1\\
    0 & -\partial_0
  \end{pmatrix}.
  \label{Delta-old-matrix}
\end{equation}

The determinant is represented by fermionic ghosts.  It is convenient to package the two antighost
components as the surviving components of a symmetric tensor,
\begin{equation}
  b^0\equiv b_{00},\qquad b^1\equiv b_{01}=b_{10}.
\end{equation}
Unlike the tensile \(b_{ab}\), this BMS antighost is not imposed to be traceless.  The reason is
geometric: the null gauge \(V^a=(1,0)\) is not a conformal metric gauge, and the BMS
transformations do not preserve the ordinary traceless condition on a two-tensor.
With \(c^a=(c^0,c^1)\), the gauge-fixed action is
\begin{equation}
  S_{\mathrm{ILST}+bc}
  =
  \frac{1}{2\pi}\int\dd^2\sigma\,
  \left[
  \dot X^2
  +i\left(
  c^0\partial_0b^0-c^1\partial_1b^0+2c^1\partial_0b^1
  \right)
  \right].
  \label{old-bc-action}
\end{equation}
This is the BMS \(bc\)-ghost system.  Its equations of motion are
\begin{align}
  \ddot X^\mu&=0,
  &
  \partial_0c^0-\partial_1c^1&=0,
  &
  \partial_0c^1&=0,
  \label{old-eom1}
  \\
  \partial_0b^0&=0,
  &
  \frac12\partial_1b^0-\partial_0b^1&=0 .
  \label{old-eom2}
\end{align}
The corresponding mode expansions are
\begin{align}
  X^\mu
  &=
  x^\mu+\frac12p^\mu\tau
  +\frac{i}{2}\sum_{n\neq0}\frac{1}{n}
  \left(A^\mu_n-in\tau B^\mu_n\right)e^{-in\sigma},
  \\
  c^1
  &=
  \sum_n c_ne^{-in\sigma},
  &
  c^0
  &=
  \sum_n(\tilde c_n-in\tau c_n)e^{-in\sigma},
  \\
  b^0
  &=
  \sum_n\tilde b_ne^{-in\sigma},
  &
  b^1
  &=
  \frac12\sum_n(b_n-in\tau\tilde b_n)e^{-in\sigma}.
\end{align}
Canonical quantization gives
\begin{equation}
  [A^\mu_m,B^\nu_n]=2m\,\eta^{\mu\nu}\delta_{m+n,0},
  \qquad
  \{\tilde b_n,\tilde c_m\}=\delta_{m+n,0},
  \qquad
  \{b_n,c_m\}=\delta_{m+n,0}.
\end{equation}
The BMS stress-tensor components read
\begin{align}
  2\pi T_1
  &=
  2\dot X\cdot X'
  -i\left[
  2\partial_1(c^0)b_{00}+c^0\partial_1b_{00}
  +4\partial_1(c^1)b_{01}+2c^1\partial_1b_{01}
  \right],
  \\
  2\pi T_2
  &=
  \dot X^2
  -i\left[
  2\partial_1(c^1)b_{00}+c^1\partial_1b_{00}
  \right].
\end{align}
The corresponding modes
\begin{equation}
  L_n=\int\dd\sigma\,(T_1+in\tau T_2)e^{in\sigma},
  \qquad
  M_n=\int\dd\sigma\,T_2e^{in\sigma}
\end{equation}
generate the BMS algebra.  In the flipped vacuum the anomaly cancellation in this older two-ghost
system reproduces \(D=26\) \cite{Chen:2023esw}, but the point of the following sections is that this is not the complete
gauge-fixed ghost theory once the Carroll-Weyl scaling is kept as a local gauge symmetry.

\section{Carroll-Weyl gauged null string}
\label{sec:cw}

\subsection{The two Carroll-Weyl scalings}

We briefly review the two Carroll-Weyl scalings of a null-string worldsheet, following
\cite{SheikhJabbari:2026I}.  In an ordinary Lorentzian or Euclidean two-dimensional
geometry, the metric $\gamma_{ab}$ admits the usual Weyl rescaling
\begin{equation}
    \gamma_{ab}\ \rightarrow\ e^{2\phi(\sigma)}\gamma_{ab}.
\end{equation}
This rescaling changes the local scale of the metric while leaving the coordinate system
unchanged.  The important point emphasized in \cite{SheikhJabbari:2026I} is that the
Carrollian worldsheet of a null string has a richer geometric structure.  Instead of a
non-degenerate metric, the intrinsic data are described by the Carrollian triple
\begin{equation}
    \left(\ell_a,\ v^a,\ n_a\right),
\end{equation}
where $v^a$ is the kernel vector, $n_a$ is the clock one-form normalized by
\begin{equation}
    v^a n_a = 1,
\end{equation}
and $\ell_a$ is the spatial one-form defining the degenerate spatial metric
\begin{equation}
    h_{ab}=\ell_a \ell_b .
\end{equation}
The associated two-dimensional volume form is
\begin{equation}
    \epsilon_{ab}=n_a\ell_b-n_b\ell_a .
\end{equation}

Because the Carrollian structure separates the temporal and spatial directions, one can
rescale the clock sector and the spatial sector independently.  Thus one has
\begin{equation}
    n_a \ \rightarrow\ e^{\chi_t(\sigma)} n_a ,
    \qquad
    v^a \ \rightarrow\ e^{-\chi_t(\sigma)} v^a ,
\end{equation}
together with
\begin{equation}
    \ell_a \ \rightarrow\ e^{\chi_s(\sigma)} \ell_a .
\end{equation}
The normalization condition $v^a n_a=1$ is preserved by the first transformation.  The
derived geometric objects then transform as
\begin{equation}
    h_{ab}=\ell_a\ell_b
    \ \rightarrow\
    e^{2\chi_s} h_{ab},
\end{equation}
and
\begin{equation}
    \epsilon_{ab}=n_a\ell_b-n_b\ell_a
    \ \rightarrow\
    e^{\chi_t+\chi_s}\epsilon_{ab}.
\end{equation}

It is useful to decompose these two independent scalings into two special combinations.
The first is the symmetric combination
\begin{equation}
    \phi:=\chi_t=\chi_s .
\end{equation}
This is the Carrollian analogue of the ordinary Weyl rescaling, because both the volume
form and the spatial metric scale in the standard way
\begin{equation}
    n_a \ \rightarrow\ e^\phi n_a ,
    \qquad
    v^a \ \rightarrow\ e^{-\phi}v^a ,
    \qquad
    \ell_a \ \rightarrow\ e^\phi \ell_a ,
\end{equation}
and therefore
\begin{equation}
    \epsilon_{ab}\ \rightarrow\ e^{2\phi}\epsilon_{ab},
    \qquad
    h_{ab}\ \rightarrow\ e^{2\phi}h_{ab}.
\end{equation}
For this reason, this transformation may be called the \emph{volume-modulating}
Carroll-Weyl scaling.

In two dimensions one may choose a basis in which the degenerate metric takes the form
\begin{equation}
    h_{ab}=
    \begin{pmatrix}
    0 & 0 \\
    0 & h
    \end{pmatrix}.
\end{equation}
Then the vector density
\begin{equation}
    \mathcal V^a:=\sqrt{h}\,v^a
\end{equation}
is invariant under the $\phi$-scaling.  This object is the Carrollian analogue of
$\sqrt{-\gamma}\gamma^{ab}$ in the Polyakov formulation.  Under worldsheet
diffeomorphisms it transforms as a vector density of weight $1/2$,
\begin{equation}
    \delta_\xi \mathcal V^a
    =
    \xi^b\partial_b \mathcal V^a
    -
    \mathcal V^b\partial_b\xi^a
    +
    \frac12(\partial_b\xi^b)\mathcal V^a .
\end{equation}
This should be compared with the transformation of the ordinary Polyakov combination
$\sqrt{-\gamma}\gamma^{ab}$,
\begin{equation}
    \delta_\xi\left(\sqrt{-\gamma}\gamma^{ab}\right)
    =
    \sqrt{-\gamma}
    \left(
    \nabla^a\xi^b
    +
    \nabla^b\xi^a
    +
    (\nabla_c\xi^c)\gamma^{ab}
    \right),
\end{equation}
where $\xi_a=\gamma_{ab}\xi^b$ and $\nabla_a$ is the covariant derivative associated with
$\gamma_{ab}$.

The second and more intrinsically Carrollian combination is obtained by taking
\begin{equation}
    \chi:=\chi_t=-\chi_s .
\end{equation}
This gives
\begin{equation}
    n_a \ \rightarrow\ e^\chi n_a ,
    \qquad
    v^a \ \rightarrow\ e^{-\chi}v^a ,
    \qquad
    \ell_a \ \rightarrow\ e^{-\chi}\ell_a .
\end{equation}
Consequently,
\begin{equation}
    \mathcal V^a \ \rightarrow\ e^{-2\chi}\mathcal V^a ,
    \qquad
    h_{ab}\ \rightarrow\ e^{-2\chi}h_{ab},
\end{equation}
whereas the volume form is left invariant
\begin{equation}
    \epsilon_{ab}
    \ \rightarrow\
    e^{\chi-\chi}\epsilon_{ab}
    =
    \epsilon_{ab}.
\end{equation}
This is therefore a \emph{volume-preserving Carroll-Weyl scaling}.  It has no direct
counterpart in an ordinary Lorentzian worldsheet geometry, where the metric and the
volume form are tied together by non-degeneracy.  The existence of this second scaling is
one of the central geometric observations of \cite{SheikhJabbari:2026I}.  It is precisely
this extra Carrollian scaling option that motivates the gauged Carroll-Weyl formulation
of the null string and ultimately leads to the additional gauge structure absent in the
standard tensile-string Polyakov description.

\subsection{Null string action and the constraints}
The Carrollian worldsheet admits a second Weyl-type scaling not present in Lorentzian geometry
\cite{SheikhJabbari:2026I}.  In the normalization used in this paper the volume-preserving
Carroll-Weyl transformation acts as
\begin{equation}
  \delta_\chi X^\mu=\chi X^\mu,\qquad
  \delta_\chi V^a=-\chi V^a .
  \label{cw-scaling}
\end{equation}
The ILST action is not invariant under \eqref{cw-scaling} for generic \(\chi(\tau,\sigma)\).  The
standard remedy is to introduce a Carroll-Weyl connection \(W_a\) and define
\begin{equation}
  D_aX^\mu=\partial_aX^\mu+W_aX^\mu,
  \qquad
  \delta_\chi W_a=-\partial_a\chi .
  \label{covD}
\end{equation}
Then
\begin{equation}
  \delta_\chi(D_aX^\mu)=\chi D_aX^\mu,
\end{equation}
and the gauged null-string action
\begin{equation}
  S_{\CW}[X,V,W]
  =
  \frac{1}{2\pi}\int\dd^2\sigma\,
  V^aV^bD_aX\cdot D_bX
  =
  \frac{1}{2\pi}\int\dd^2\sigma\,(\cD X)^2
  \label{cw-action}
\end{equation}
is invariant under local Carroll-Weyl transformations.  Here
\begin{equation}
  \cD X^\mu\equiv V^aD_aX^\mu
  =
  V^a\partial_aX^\mu+\cW X^\mu,
  \qquad
  \cW\equiv V^aW_a .
  \label{mathcalD}
\end{equation}
Only the contraction \(\cW\) enters the action.

The combined infinitesimal diffeomorphism plus Carroll-Weyl transformations are
\begin{align}
  \delta_\eta X^\mu
  &=
  \xi^a\partial_aX^\mu+\chi X^\mu,
  \label{full-delta-X}
  \\
  \delta_\eta V^a
  &=
  -V^b\partial_b\xi^a+\xi^b\partial_bV^a
  +\frac12V^a\partial_b\xi^b-\chi V^a,
  \label{full-delta-V}
  \\
  \delta_\eta W_a
  &=
  \xi^b\partial_bW_a+W_b\partial_a\xi^b-\partial_a\chi .
  \label{full-delta-W}
\end{align}
Under \(\chi\) alone,
\begin{equation}
  \delta_\chi(\cD X^\mu)
  =
  \delta_\chi(V^aD_aX^\mu)
  =
  -\chi V^aD_aX^\mu+\chi V^aD_aX^\mu=0,
\end{equation}
so the Carroll-Weyl invariance of \eqref{cw-action} is manifest.  Under diffeomorphisms,
\(\cD X^\mu\) transforms as a density of weight \(1/2\), so \((\cD X)^2\dd^2\sigma\) is
diffeomorphism invariant.

The equations obtained by varying the auxiliary gauge fields are constraints.  Define the canonical
momentum density
\begin{equation}
  P_\mu\equiv \cD X_\mu .
  \label{Pdef}
\end{equation}
Varying \(V^a\) gives
\begin{equation}
  \delta_V S_{\CW}
  =
  \frac{1}{\pi}\int\dd^2\sigma\,
  \delta V^a\,D_aX\cdot P,
\end{equation}
so
\begin{equation}
  P\cdot D_aX=0,\qquad a=0,1.
  \label{VDconstraint}
\end{equation}
Varying \(W_a\) gives
\begin{equation}
  \delta_W S_{\CW}
  =
  \frac{1}{\pi}\int\dd^2\sigma\,
  P\cdot X\,V^a\delta W_a,
\end{equation}
so the Carroll-Weyl constraint is
\begin{equation}
  C_3\equiv P\cdot X=0.
  \label{C3}
\end{equation}
After using \(P\cdot D_aX=P\cdot\partial_aX+\cW P\cdot X\), the full constraint set may be written as
\begin{equation}
  C_a\equiv P\cdot\partial_aX=0,
  \qquad
  C_3\equiv P\cdot X=0 .
\end{equation}
In the gauge \(V^a=(1,0)\), \(\cW=0\), one has \(P_\mu=\dot X_\mu\), and the three independent
constraints are
\begin{equation}
  C_1=P^2=0,\qquad
  C_2=P\cdot X'=0,\qquad
  C_3=P\cdot X=0 .
  \label{three-constraints}
\end{equation}
The third one is the new ingredient relative to the older ILST path integral.

\section{Revised path integral and the \texorpdfstring{\(bcs\)}{bcs}-ghost determinant}
\label{sec:fp-bcs}

\subsection{Gauge conditions}

The unfixed Carroll-Weyl covariant path integral is formally
\begin{equation}
  Z_{\CW}
  =
  \int
  \frac{DX\,DV\,DW}{\Vol(\Diff\ltimes\CW)}
  \exp\{iS_{\CW}[X,V,W]\}.
  \label{unfixed-cw-path-integral}
\end{equation}
The semidirect product notation emphasizes that Carroll-Weyl scalings do not form an inert spectator
factor: diffeomorphisms act on the local scaling parameter by pullback, and the constraint generated by
the scaling has non-trivial brackets with the BMS constraints.  A correct Faddeev-Popov treatment must
therefore divide by \(\Diff\ltimes\CW\), not only by the diffeomorphism subgroup.

The gauge group is now generated by
\begin{equation}
  \eta^A=(\epsilon^0,\epsilon^1,\lambda),
\end{equation}
where \(\lambda\) is the Carroll-Weyl parameter.  We fix the gauge by imposing
\begin{equation}
  G^0=V^0-1=0,\qquad
  G^1=V^1=0,\qquad
  G^s=\cW=V^aW_a=0 .
  \label{gauge-conditions}
\end{equation}
The last condition is the diffeomorphism-invariant Carroll-Weyl gauge used in
\cite{SheikhJabbari:2026I,SheikhJabbari:2026II}.  In the temporal gauge it is simply \(W_0=0\).
The component of \(W_a\) orthogonal to \(V^a\) does not enter \eqref{cw-action}; it only contributes an
overall decoupled volume factor and will be suppressed below.

The gauge-preserving equations already reveal the physical change.  Setting
\(\delta G^0=\delta G^1=\delta G^s=0\) gives
\begin{equation}
  \partial_0\epsilon^1=0,\qquad
  \partial_0\lambda=0,\qquad
  \partial_0\epsilon^0=\partial_1\epsilon^1-2\lambda .
  \label{residual-cw-eq}
\end{equation}
Thus the residual transformations are labelled by three functions of the spatial circle
\begin{equation}
  \epsilon^1=f(\sigma),\qquad
  \lambda=\chi(\sigma),\qquad
  \epsilon^0=\tau\big(f'(\sigma)-2\chi(\sigma)\big)+g(\sigma).
  \label{residual-cw}
\end{equation}
The old BMS residual symmetry is recovered by setting \(\chi=0\).  The extra function \(\chi\)
is precisely the residual Carroll-Weyl scaling.  This is why the ghost equation below will contain
\(\partial_0c^0-\partial_1c^1+2s=0\), rather than the old BMS equation
\(\partial_0c^0-\partial_1c^1=0\).

Equivalently, the three gauge functions change linearly as
\begin{equation}
  \begin{pmatrix}
    \delta G^0\\[2pt]
    \delta G^1\\[2pt]
    \delta G^s
  \end{pmatrix}
  =
  \begin{pmatrix}
    -\frac12\partial_0\epsilon^0+\frac12\partial_1\epsilon^1-\lambda\\[2pt]
    -\partial_0\epsilon^1\\[2pt]
    -\partial_0\lambda
  \end{pmatrix}.
  \label{linearized-gauge-functions}
\end{equation}
This already displays the three independent infinitesimal gauge directions that must be divided out in
the path integral.  The first row is especially important: the same gauge condition \(V^0=1\) is moved
both by time reparametrizations and by Carroll-Weyl rescalings, which is the origin of the off-diagonal
\(-1\) entry in the Faddeev-Popov matrix.

The Faddeev-Popov identity is
\begin{equation}
  1
  =
  \int D\epsilon^0D\epsilon^1D\lambda\,
  \delta(G^{0,\eta})\delta(G^{1,\eta})\delta(G^{s,\eta})
  \det \cM ,
  \label{fp-identity-new}
\end{equation}
where
\begin{equation}
  \cM^A{}_B(\sigma,\sigma')
  =
  \left.
  \frac{\delta G^{A,\eta}(\sigma)}
  {\delta\eta^B(\sigma')}
  \right|_{\eta=0},
  \qquad
  A,B\in\{0,1,s\}.
\end{equation}
Here \(G^{A,\eta}\) denotes the gauge-fixing function \(G^A\) evaluated after a gauge transformation generated by \(\eta^B=(\epsilon^0,\epsilon^1,\lambda)\). Infinitesimally,
\begin{equation}
G^{A,\eta}
=
G^A+\delta_\eta G^A,
\qquad
\delta_\eta G^A
=
\cM^A{}_B\,\eta^B\equiv\delta G^A.
\end{equation}
The functional delta functions \(\delta(G^{A,\eta})\) enforce the gauge conditions by selecting the representative on each gauge orbit satisfying \(G^A=0\).

After inserting \eqref{fp-identity-new}, the path integral on the gauge slice is
\begin{equation}
  Z_{\CW}
  =
  \int DX\,\det\cM\,
  \exp\!\left\{
  \frac{i}{2\pi}\int\dd^2\sigma\,\dot X^2
  \right\},
  \label{cw-sliced-path-integral}
\end{equation}
up to the decoupled volume coming from the component of \(W_a\) orthogonal to \(V^a\).

\subsection{The three-row Faddeev-Popov operator}

Using \eqref{full-delta-V}, evaluated on \(V^a=(1,0)\), we find
\begin{align}
  \delta V^0
  &=
  -\frac12\partial_0\epsilon^0+\frac12\partial_1\epsilon^1-\lambda,
  \label{deltaV0-new}
  \\
  \delta V^1
  &=
  -\partial_0\epsilon^1 .
  \label{deltaV1-new}
\end{align}
The third gauge condition is slightly more instructive.  Since \(\cW=V^aW_a\),
\begin{equation}
  \delta\cW=(\delta V^a)W_a+V^a\delta W_a .
\end{equation}
On the gauge slice \(V^a=(1,0)\), \(V^aW_a=0\), this gives
\begin{align}
  \delta\cW
  &=
  (\delta V^1)W_1+\delta W_0
  \notag\\
  &=
  (-\partial_0\epsilon^1)W_1+
  \left(W_1\partial_0\epsilon^1-\partial_0\lambda\right)
  \notag\\
  &=
  -\partial_0\lambda .
  \label{deltaWcal}
\end{align}
The dependence on the decoupled spatial component \(W_1\) cancels, as it must.

We now extract the operator \(\cM\) directly.  By definition, the Faddeev-Popov operator is the
linear map from gauge parameters to gauge-condition variations
\begin{equation}
  \delta G^A
  =
  \cM^A{}_B\eta^B,
  \qquad
  \eta^B=(\epsilon^0,\epsilon^1,\lambda),
  \qquad
  G^A=(G^0,G^1,G^s).
  \label{deltaG-Meta}
\end{equation}
In components this says
\begin{align}
  \delta G^0
  &=
  \cM^0{}_0\epsilon^0+\cM^0{}_1\epsilon^1+\cM^0{}_s\lambda,
  \label{component-row-0}
  \\
  \delta G^1
  &=
  \cM^1{}_0\epsilon^0+\cM^1{}_1\epsilon^1+\cM^1{}_s\lambda,
  \label{component-row-1}
  \\
  \delta G^s
  &=
  \cM^s{}_0\epsilon^0+\cM^s{}_1\epsilon^1+\cM^s{}_s\lambda .
  \label{component-row-s}
\end{align}
Comparing \eqref{component-row-0}--\eqref{component-row-s} with
\eqref{linearized-gauge-functions}, or equivalently with
\eqref{deltaV0-new}, \eqref{deltaV1-new}, and \eqref{deltaWcal}, gives row by row
\begin{align}
  \delta G^0
  &=
  -\frac12\partial_0\epsilon^0
  +\frac12\partial_1\epsilon^1
  -\lambda
  \notag\\
  &\Rightarrow\quad
  \cM^0{}_0=-\frac12\partial_0,\qquad
  \cM^0{}_1=\frac12\partial_1,\qquad
  \cM^0{}_s=-1,
  \label{M-row-zero}
  \\
  \delta G^1
  &=
  -\partial_0\epsilon^1
  \notag\\
  &\Rightarrow\quad
  \cM^1{}_0=0,\qquad
  \cM^1{}_1=-\partial_0,\qquad
  \cM^1{}_s=0,
  \label{M-row-one}
  \\
  \delta G^s
  &=
  -\partial_0\lambda
  \notag\\
  &\Rightarrow\quad
  \cM^s{}_0=0,\qquad
  \cM^s{}_1=0,\qquad
  \cM^s{}_s=-\partial_0 .
  \label{M-row-s}
\end{align}
Thus, ordered as \((G^0,G^1,G^s)\) and \((\epsilon^0,\epsilon^1,\lambda)\), the Faddeev-Popov
operator is
\begin{equation}
  \cM
  =
  \begin{pmatrix}
    -\frac12\partial_0 & \frac12\partial_1 & -1\\
    0 & -\partial_0 & 0\\
    0 & 0 & -\partial_0
  \end{pmatrix}.
  \label{Mmatrix}
\end{equation}
The old BMS \(bc\) operator \eqref{Delta-old-matrix} is the upper-left block.  The third column and
third row are the path-integral imprint of Carroll-Weyl scaling.
For example, \(\cM^0{}_s=-1\) comes from the Carroll-Weyl variation
\(\delta_\lambda(V^0-1)=-\lambda\), while \(\cM^s{}_s=-\partial_0\) comes from
\(\delta_\lambda(V^aW_a)=-\partial_0\lambda\).  These two entries are what distinguish the
Carroll-Weyl covariant determinant from the old two-row BMS determinant.  A longer derivation of each
row is given in appendix \ref{app:fp-details}.
Since \(\cM\) is triangular, its formal determinant is
\begin{equation}
  \det\cM
  =
  \det\!\left(-\frac12\partial_0\right)
  \det(-\partial_0)
  \det(-\partial_0).
  \label{triangular-det}
\end{equation}
Although the determinant of the triangular Faddeev-Popov operator $\cM$ depends only on its diagonal entries, the ghost action  depends on the full operator. The off-diagonal terms \(\cM_{01}\) and \(\cM_{0s}\) encode how a single gauge orbit moves through the gauge slice, namely that \(G^{0}\) is affected by both spatial diffeomorphisms and Carroll-Weyl transformations. Upon exponentiation, this gauge-orbit mixing is precisely what generates the local BMS ghost couplings \(c^{1}\partial_{1} b^{0}\) and the new Carroll-Weyl mixing \(s\,b^{0}\), to be discussed in the next subsection.

\subsection{Exponentiation and the \texorpdfstring{\(bcs\)}{bcs}-ghost action}

Introduce fermionic ghosts
\begin{equation}
  c^A=(c^0,c^1,s),
\end{equation}
where \(s\) is the Carroll-Weyl ghost, and antighosts
\begin{equation}
  b_A=(2b^0,2b^1,b^s).
\end{equation}
The factors of \(2\) in the first two antighosts are the same normalization that converts the old
two-row determinant into the standard BMS \(bc\)-ghost action \cite{Chen:2023esw}.  The determinant is exponentiated by
the Grassmann Gaussian identity.  For a finite-dimensional matrix \(M^A{}_B\),
\begin{equation}
  \det M
  =
  \int \prod_A db_A\,dc^A\,
  \exp\!\left[-b_AM^A{}_Bc^B\right],
  \label{finite-grassmann-det}
\end{equation}
up to an overall normalization fixed by the ordering of the Berezin measure. The only ambiguity is the overall sign fixed by the orientation of the Berezin measure; since this sign is field independent, it is absorbed into the normalization of the path integral. A short proof of this Grassmann identity, including the measure convention, is given in Appendix~\ref{app:grassmann-det}. 

In a Lorentzian worldsheet
path integral we want the full gauge-fixed integrand to be of the form \(\exp(iS)\).  Therefore we write
\begin{equation}
  \exp\!\left[-\int\dd^2\sigma\,b_A\cM^A{}_Bc^B\right]
  =
  \exp\!\left[
  i\left(
  -i\int\dd^2\sigma\,b_A\cM^A{}_Bc^B
  \right)
  \right].
  \label{lorentzian-ghost-phase}
\end{equation}
Equivalently, the factors of \(i\) may be absorbed into a constant phase of the Grassmann determinant,
which is independent of the dynamical fields.  With this convention,
\begin{equation}
  \det\cM
  =
  \int Db^0Db^1Db^sDc^0Dc^1Ds\,
  \exp\{iS_{bcs}\},
  \label{detM-bcs-convention}
\end{equation}
and the ghost action is obtained from
\begin{equation}
  iS_{bcs}
  \sim
  -i\int \dd^2\sigma\,
  b_A\cM^A{}_Bc^B .
  \label{iSbcs-from-det}
\end{equation}
Here \(\sim\) means equality up to field-independent determinant normalizations and the common
worldsheet factor \(1/(2\pi)\), which is restored in \eqref{bcs-action}.  Equation
\eqref{iSbcs-from-det} is simply the Lorentzian rewriting of the Grassmann determinant
\eqref{finite-grassmann-det}.
Substituting \eqref{Mmatrix} gives, row by row,
\begin{align}
  -b_A\cM^A{}_Bc^B
  &=
  -2b^0\left(
  -\frac12\partial_0c^0+\frac12\partial_1c^1-s
  \right)
  -2b^1(-\partial_0c^1)
  -b^s(-\partial_0s)
  \notag\\
  &=
  b^0\partial_0c^0-b^0\partial_1c^1+2b^0s
  +2b^1\partial_0c^1+b^s\partial_0s .
  \label{row-by-row-ghost}
\end{align}
Thus the pre-integration-by-parts form is
\begin{align}
  -b_A\cM^A{}_Bc^B
  &=
  b^0\partial_0c^0-b^0\partial_1c^1+2b^0s
  +2b^1\partial_0c^1+b^s\partial_0s .
  \label{preparts}
\end{align}
Here the cylinder is the closed-string worldsheet
\begin{equation}
  \Sigma_{\mathrm{cyl}}
  =
  [\tau_i,\tau_f]\times S^1,
  \qquad
  \sigma\sim\sigma+2\pi,
  \qquad
  \int_{\Sigma_{\mathrm{cyl}}}\dd^2\sigma
  =
  \int_{\tau_i}^{\tau_f}\dd\tau\int_0^{2\pi}\dd\sigma .
  \label{worldsheet-cylinder}
\end{equation}
All fields are periodic around \(S^1\).  In deriving the local action we either fix the variations at
\(\tau_i,\tau_f\) or take fields with compact support in time, so total \(\tau\)-derivatives and total
\(\sigma\)-derivatives do not contribute.  Because the ghosts and antighosts are Grassmann odd,
\(\int_{\Sigma_{\mathrm{cyl}}} b\,\partial_a c=\int_{\Sigma_{\mathrm{cyl}}}c\,\partial_a b\) for two
odd fields \(b,c\).  Therefore integration by parts gives
\begin{align}
  \int\dd^2\sigma\,b^0\partial_0c^0
  &=
  \int\dd^2\sigma\,c^0\partial_0b^0,
  \\
  -\int\dd^2\sigma\,b^0\partial_1c^1
  &=
  -\int\dd^2\sigma\,c^1\partial_1b^0,
  \\
  2\int\dd^2\sigma\,b^1\partial_0c^1
  &=
  2\int\dd^2\sigma\,c^1\partial_0b^1,
  \\
  \int\dd^2\sigma\,b^s\partial_0s
  &=
  \int\dd^2\sigma\,s\partial_0b^s,
  \\
  2b^0s&=-2sb^0 .
\end{align}
Thus the gauge-fixed matter plus ghost action is
\begin{equation}
  S_{\mathrm{gf}}
  =
  \frac{1}{2\pi}\int\dd^2\sigma\,
  \left[
  \dot X^2
  +i\left(
  c^0\partial_0b^0
  -c^1\partial_1b^0
  +2c^1\partial_0b^1
  +s\partial_0b^s
  -2sb^0
  \right)
  \right].
  \label{bcs-action}
\end{equation}
Equation \eqref{bcs-action} is the central result of the path-integral revision.  The first three ghost
terms are the BMS \(bc\) system.  The last two terms are forced by the Carroll-Weyl determinant.
The kinetic term \(s\partial_0b^s\) comes from fixing \(\cW=0\), and the algebraic mixing
\(-2sb^0\) comes from the fact that \(V^0\) itself scales under Carroll-Weyl transformations.
Appendix \ref{app:ghost-action-details} gives a more detailed component derivation of
\eqref{bcs-action}, including the Grassmann signs and the role of the off-diagonal entries of
\(\cM_{\CW}\).

\section{Equations of motion and mode expansion of the \texorpdfstring{\(bcs\)}{bcs} system}
\label{sec:modes}

We now derive the free equations of motion of the \(bcs\) system directly from \eqref{bcs-action}.
The matter part is
\begin{equation}
  S_X^{\mathrm{gf}}
  =
  \frac{1}{2\pi}\int\dd^2\sigma\,\dot X^2 .
\end{equation}
Its variation is
\begin{equation}
  \delta S_X^{\mathrm{gf}}
  =
  \frac{1}{\pi}\int\dd^2\sigma\,\dot X\cdot\partial_0\delta X
  =
  -\frac{1}{\pi}\int\dd^2\sigma\,\ddot X\cdot\delta X,
\end{equation}
where the cylinder boundary conditions of \eqref{worldsheet-cylinder} have been used.  Hence the
matter equation is
\begin{equation}
  \ddot X^\mu=0,
\end{equation}
as in the ordinary ILST gauge-fixed action.

For the ghosts, write
\begin{equation}
  \mathcal L_{bcs}
  =
  i\left(
  c^0\partial_0b^0
  -c^1\partial_1b^0
  +2c^1\partial_0b^1
  +s\partial_0b^s
  -2sb^0
  \right).
  \label{bcs-lagrangian}
\end{equation}
Varying the ghosts \(c^0,c^1,s\), which appear without derivatives, gives the antighost equations
\begin{align}
  \partial_0b^0&=0,
  &
  \frac12\partial_1b^0-\partial_0b^1&=0,
  &
  \partial_0b^s-2b^0&=0,
  \label{beom}
\end{align}
respectively.  Varying the antighosts \(b^0,b^1,b^s\) and integrating by parts on
\(\Sigma_{\mathrm{cyl}}\) gives the ghost equations
\begin{align}
  \partial_0c^1&=0,
  &
  \partial_0s&=0,
  &
  \partial_0c^0-\partial_1c^1+2s&=0 .
  \label{ceom}
\end{align}
The last equation is precisely the ghost version of the gauge-preserving condition
\(\delta V^0=0\) in \eqref{deltaV0-new}; the new \(s\) term is not optional.
Appendix \ref{app:modes-details} gives the same variation with every sign displayed.

We solve these equations on the same cylinder.  Since \(\ddot X^\mu=0\), the embedding is linear in
\(\tau\).  Expanding the two arbitrary \(\sigma\)-dependent functions in Fourier modes gives
\begin{align}
  X^\mu
  &=
  x^\mu+\frac12p^\mu\tau
  +\frac{i}{2}\sum_{n\neq0}\frac1n
  \left(A^\mu_n-in\tau B^\mu_n\right)e^{-in\sigma},
  \label{X-mode-bcs}
\end{align}
where the normalization has been chosen so that the canonical brackets take the standard tensionless
form.  The equations \(\partial_0c^1=0\) and \(\partial_0s=0\) imply
\begin{align}
  c^1
  &=
  \sum_n c_ne^{-in\sigma},
  &
  s
  &=
  \sum_n s_ne^{-in\sigma}.
\end{align}
The remaining ghost equation reads
\begin{equation}
  \partial_0c^0
  =
  \partial_1c^1-2s
  =
  \sum_n(-in c_n-2s_n)e^{-in\sigma}.
\end{equation}
Integrating once in \(\tau\) gives
\begin{align}
  c^0
  &=
  \sum_n\left(\tilde c_n-in\tau c_n-2\tau s_n\right)e^{-in\sigma}.
  \label{c0-new-mode}
\end{align}
Similarly, \(\partial_0b^0=0\) gives
\begin{equation}
  b^0
  =
  \sum_n\tilde b_ne^{-in\sigma}.
\end{equation}
The equation \(\partial_0b^1=\frac12\partial_1b^0\) gives
\begin{align}
  b^1
  &=
  \frac12\sum_n(b_n-in\tau\tilde b_n)e^{-in\sigma},
\end{align}
and \(\partial_0b^s=2b^0\) gives
\begin{align}
  b^s
  &=
  \sum_n(r_n+2\tau\tilde b_n)e^{-in\sigma}.
  \label{bs-mode}
\end{align}
Collecting the ghost modes,
\begin{align}
  c^1
  &=
  \sum_n c_ne^{-in\sigma},
  \\
  s
  &=
  \sum_n s_ne^{-in\sigma},
  \\
  c^0
  &=
  \sum_n\left(\tilde c_n-in\tau c_n-2\tau s_n\right)e^{-in\sigma},
  \\
  b^0
  &=
  \sum_n\tilde b_ne^{-in\sigma},
  \\
  b^1
  &=
  \frac12\sum_n(b_n-in\tau\tilde b_n)e^{-in\sigma},
  \\
  b^s
  &=
  \sum_n(r_n+2\tau\tilde b_n)e^{-in\sigma}.
  \label{all-bcs-modes}
\end{align}
The new terms in \(c^0\) and \(b^s\) are paired: \(s\) deforms the temporal diffeomorphism ghost, and
\(b^0\) sources the Carroll-Weyl antighost.

Canonical quantization gives
\begin{equation}
  [x^\mu,p^\nu]=i\eta^{\mu\nu},
  \qquad
  [A^\mu_m,B^\nu_n]=2m\,\eta^{\mu\nu}\delta_{m+n,0},
\end{equation}
and
\begin{equation}
  \{b_n,c_m\}=\delta_{m+n,0},
  \qquad
  \{\tilde b_n,\tilde c_m\}=\delta_{m+n,0},
  \qquad
  \{r_n,s_m\}=\delta_{m+n,0}.
  \label{bcs-anticommutators}
\end{equation}
The algebraic term \(-2sb^0\) does not change the equal-time brackets because it contains no
\(\tau\)-derivative.  It does, however, change the Hamiltonian evolution: \(s\) enters the solution for
\(c^0\), and \(b^0\) enters the solution for \(b^s\).  This is the canonical trace of the same
off-diagonal Faddeev-Popov entry that produced the mixing term in the action. For a detailed derivation of the oscillator algebra \eqref{bcs-anticommutators}, see Appendix~\ref{app:oscillatoralgebra}.

\section{Extended BMS constraints and BRST structure}
\label{sec:extended}

The gauge-fixed action \eqref{bcs-action} still carries the constraints generated by the three residual
gauge symmetries.  For the matter fields the constraints are \eqref{three-constraints}.  Their smeared
forms are
\begin{equation}
  C_1[f]=\int\dd\sigma\,f(\sigma)C_1(\sigma),
  \quad
  C_2[g]=\int\dd\sigma\,g(\sigma)C_2(\sigma),
  \quad
  C_3[k]=\int\dd\sigma\,k(\sigma)C_3(\sigma).
\end{equation}
Using the canonical Poisson brackets of \(X^\mu\) and \(P_\mu\), one obtains the closed algebra
\cite{SheikhJabbari:2026II}
\begin{equation}
  \{X^\mu(\sigma),P_\nu(\sigma')\}
  =
  \delta^\mu{}_\nu\,\delta(\sigma-\sigma').
  \label{canonical-pb}
\end{equation}
The genuinely new bracket can be derived in one line:
\begin{align}
  \{C_1[f],C_3[g]\}
  &=
  \int\dd\sigma\,\dd\sigma'\,
  f(\sigma)g(\sigma')
  \{P^2(\sigma),P(\sigma')\cdot X(\sigma')\}
  \notag\\
  &=
  -2\int\dd\sigma\,f(\sigma)g(\sigma)P^2(\sigma)
  =
  -2C_1[fg].
  \label{C1C3-derivation}
\end{align}
Thus \(C_3\) is not central: it rescales \(C_1\), exactly as expected from
\(\delta_\lambda P_\mu=-\lambda P_\mu\), \(\delta_\lambda X^\mu=\lambda X^\mu\).  The remaining
brackets follow by the same distributional manipulations and give
\begin{align}
  \{C_1[f],C_1[g]\}&=0,
  &
  \{C_1[f],C_2[g]\}&=C_1[fg'-f'g],
  \\
  \{C_1[f],C_3[g]\}&=-2C_1[fg],
  &
  \{C_2[f],C_2[g]\}&=C_2[fg'-f'g],
  \\
  \{C_2[f],C_3[g]\}&=C_3[fg'],
  &
  \{C_3[f],C_3[g]\}&=0 .
  \label{smeared-algebra}
\end{align}
With modes
\begin{equation}
  L_n=\int_0^{2\pi}\dd\sigma\,e^{-in\sigma}C_2(\sigma),
  \quad
  M_n=\int_0^{2\pi}\dd\sigma\,e^{-in\sigma}C_1(\sigma),
  \quad
  S_n=\int_0^{2\pi}\dd\sigma\,e^{-in\sigma}C_3(\sigma),
\end{equation}
this becomes the extended BMS algebra
\begin{align}
  \{L_m,L_n\}&=i(m-n)L_{m+n},
  &
  \{L_m,M_n\}&=i(m-n)M_{m+n},
  &
  \{M_m,M_n\}&=0,
  \\
  \{L_m,S_n\}&=inS_{m+n},
  &
  \{M_m,S_n\}&=-2M_{m+n},
  &
  \{S_m,S_n\}&=0 .
  \label{extended-bms}
\end{align}
The \(L,M\) subalgebra is BMS\(_3\).  The \(S_n\) modes form a weight-one current under
superrotations and implement Carroll-Weyl scaling.  Since \(\{M,S\}\neq0\), the Carroll-Weyl
constraint is not a spectator.

The BRST charge must therefore contain three ghost families.  A compact way to state the minimal
charge is to collect the constraints into \(T_A=(L,M,S)\), the ghosts into
\(\mathsf c^A=(c,\tilde c,s)\), and the antighost modes into
\(\mathsf b_A=(b,\tilde b,r)\).  If the quantum extended algebra is written as
\begin{equation}
  [T_A{}_m,T_B{}_n]
  =
  f_{AB}{}^C(m,n)\,T_C{}_{m+n}
  +K_{AB}(m)\delta_{m+n,0},
  \label{generic-extended-algebra}
\end{equation}
then the minimal BRST charge has the BFV form
\begin{equation}
  Q_{\mathrm{BRST}}
  =
  \sum_{n,A}\mathsf c^A_{-n}T_A{}_n
  -\frac12
  \sum_{m,n}
  f_{AB}{}^C(m,n):
  \mathsf c^A_{-m}\mathsf c^B_{-n}\mathsf b_{C,m+n}:
  +Q_{\mathrm{intercept}} .
  \label{BRST-BFV}
\end{equation}
Here \(K_{AB}(m)\) denotes the possible quantum central extensions of the extended BMS--Carroll-Weyl algebra, while \(Q_{\mathrm{intercept}}\) represents the normal-ordering (intercept) contributions that may be required to restore BRST nilpotency at the quantum level.

For the \(L,M\) subalgebra this reduces to the BMS BRST charge of the older \(bc\) treatment.  The
new terms are forced by the brackets involving \(S_n\), in particular the brackets corresponding to
\(\{L,S\}\) and \(\{M,S\}\) in \eqref{extended-bms}.  In a Fourier convention where the quantum
commutators are obtained from \eqref{extended-bms} by the usual replacement of Poisson brackets by
commutators, these terms contain the schematic structures
\begin{equation}
  Q_{\mathrm{BRST}}
  =
  Q_{\mathrm{BMS}}
  +\sum_n s_{-n}S_n
  +\sum_{m,n}\left[
  \alpha_{mn}:c_{-m}s_{-n}r_{m+n}:
  +\beta_{mn}:\tilde c_{-m}s_{-n}\tilde b_{m+n}:
  \right]
  +\cdots ,
  \label{BRST-new-terms}
\end{equation}
where \(\alpha_{mn}\) is determined by the weight-one transformation of \(S_n\) under \(L_m\), while
\(\beta_{mn}\) is determined by the non-trivial \(M\)-\(S\) bracket.  The ellipsis denotes the
ordinary \(cc b\) and \(c\tilde c\,\tilde b\) BMS terms and possible intercept terms.  The precise
signs in \eqref{BRST-new-terms} depend only on the Fourier and commutator convention; their
existence does not.

Nilpotency now requires
\begin{equation}
\begin{aligned}
  Q_{\mathrm{BRST}}^2=0
  \quad\Longleftrightarrow\quad
  K_{AB}=0
  \quad
  \text{for every sector of the extended algebra,}
   \\
  \text{up to the standard intercept shifts.}
\end{aligned}
  \label{nilpotency-extended}
\end{equation}
This is the Carroll-Weyl corrected version of the older BMS BRST condition.  The old two-ghost
condition checks only the \(L,M\) part of the algebra, whereas the complete path integral must also
check the \(S\)-sector and the \(M\)-\(S\) bracket.  The ghost number operator is correspondingly
enlarged to
\begin{equation}
  U_{bcs}
  =
  \sum_n:\left(c_{-n}b_n+\tilde c_{-n}\tilde b_n+s_{-n}r_n\right): .
  \label{bcs-ghost-number}
\end{equation}

The practical consequence is important.  A critical-dimension calculation based only on
\eqref{old-bc-action} is a calculation in a partially gauge-fixed theory where the Carroll-Weyl ghost has
been omitted.  In the fully gauged Carrollian theory, the anomaly problem has to be reformulated for
the matter plus \(bcs\) system.  The central extensions of the extended algebra are more constrained
than those of ordinary BMS\(_3\), and the \(s,b^s\) pair must be included before one can claim a final
quantum consistency condition.

At the level of physical states the difference is equally direct
\begin{equation}
  L_n|\Psi\rangle=0,\qquad
  M_n|\Psi\rangle=0,\qquad
  S_n|\Psi\rangle=0
  \quad(n\geq0\ \text{in a highest-weight prescription})
  \label{physical-state-conditions}
\end{equation}
with the usual qualifications about the chosen null-string vacuum.  The third condition is the quantum
version of \(P\cdot X=0\).  It removes the local target-space scale redundancy generated by
Carroll-Weyl transformations, and it is the physical reason the \(s,b^s\) ghosts cannot be ignored.

\section{Conclusions and future directions}
\label{sec:conclusion}

In this paper, we show that the path integral of the tensionless bosonic string must follow the complete Carrollian gauge orbit.  When we keep the volume-preserving Carroll-Weyl scaling as a local symmetry, the temporal gauge $V^a=(1,0)$ no longer fixes the gauge completely.  We must also fix the Carroll-Weyl connection by imposing $V^aW_a=0$.  The three gauge conditions
$$
G^A=(V^0-1,\ V^1,\ V^aW_a)
$$
then define the three-row Faddeev--Popov operator $\cM$ in \eqref{Mmatrix}.  Exponentiating this full operator gives the $bcs$-ghost action \eqref{bcs-action}.  We recover the BMS $bc$ system only after deleting the Carroll-Weyl row and column.  Thus the BMS $bc$ system describes a partially gauge-fixed theory, not the fully Carroll-Weyl covariant null string.
The central term in the new ghost action is
\begin{equation}
  S_{bcs}\supset
  -\frac{i}{\pi}\int\dd^2\sigma\,s\,b^0 .
  \nonumber
\end{equation}
This term is important because it is not a removable decoration.  It is the local path-integral memory
of the fact that \(V^0\) transforms under Carroll-Weyl scaling.  Equivalently, it is the ghost-level
reflection of the extended constraint algebra in which the Carroll-Weyl generator \(S_n\) has a
non-trivial bracket with the supertranslation generator \(M_n\).  Without this term the equations of
motion would incorrectly reduce to the old BMS ghost equations, the residual gauge algebra would be
represented incompletely, and the BRST operator would miss the ghost that enforces \(P\cdot X=0\).


The same point affects anomaly and critical-dimension calculations.  The \(D=26\)
based on the BMS \(bc\) system tests only the \(L_n,M_n\) part of the residual algebra \cite{Chen:2023esw}.  The fully
gauged theory must test the matter plus \(bcs\) realization of the extended BMS algebra.  A complete
quantum consistency analysis should compute all possible central terms in the \(L\)-, \(M\)-, and
\(S\)-sectors, determine the normal-ordering constants in the extended BRST charge, and impose nilpotency in the enlarged Hilbert space.  Only after this computation is done can one assign a final critical
dimension or intercept condition to the Carroll-Weyl covariant tensionless bosonic string.

There are several immediate directions, some of which will be developed in forthcoming work \cite{DuaryMaji:2026ToAppear}. First, one can construct the complete stress tensor of the $bcs$ system, including the improvement terms generated by the algebraic mixing $-2sb^0$, and use it to compute the quantum extended BMS algebra explicitly. Second, one can develop the corresponding BRST cohomology of physical states and vertex operators, together with the ghost-number selection rules associated with the $(s,r)$ ghost pair. Third,
one can build the corresponding null-string scattering measure.  In that problem the \(bcs\) action
is expected to control the insertions needed to divide by residual BMS-Carroll transformations, just as
the ordinary \(bc\) ghosts control conformal Killing zero modes and moduli in the tensile Polyakov
string.  Fourth, the relation to ambitwistor strings and scattering equations
\cite{Casali:2016atr,Casali:2017zkz,Mason:2013sva} can be revisited with the Carroll-Weyl ghost
included from the beginning.

Another interesting direction is the Carroll-Weyl covariant quantization of
tensionless \(p\)-branes. While critical-dimension conditions have recently
been obtained from the residual worldvolume symmetry algebra in a fixed partial gauge
\cite{Chen:2026tensionlessbranes}, it remains to understand how these results
are modified when the full Carroll-Weyl gauge symmetry and its associated
BRST structure are incorporated.

The analysis of this paper can be exteneded to tensionless superstrings, with worldsheet fermions $\psi^\mu$ and local supersymmetry parameters $\epsilon^\alpha$.  The bosonic analysis nevertheless gives a clear structural lesson: before introducing the BMS superghost sector, one must first enlarge the purely bosonic gauge sector itself.  In a Carroll-Weyl covariant superstring formulation, the standard ghost content should therefore be extended schematically as
\begin{equation}
(bc)\oplus(\beta\gamma)
\quad\longrightarrow\quad
(bcs)\oplus(\beta\gamma\delta).
\nonumber
\end{equation}
Here $\beta,\gamma$ denote the usual superghosts associated with local worldsheet supersymmetry, while the additional field $\delta$ schematically represents the super-Weyl partner required by the supersymmetric completion of Carroll-Weyl symmetry. Thus the $bcs$ system derived here should be regarded as the minimal bosonic backbone on which any future Carroll-Weyl covariant quantization of tensionless superstrings must be built.



\acknowledgments
We are grateful to M. M. Sheikh-Jabbari for clarifications on the earlier version of the paper, and for useful discussions. We thank Hossein Yavartanoo for clarifications on the earlier version of the paper. We thank Zezhou Hu and Ulf Lindstr\"om for useful comments. SD is supported by the Shuimu Tsinghua Scholar Program of Tsinghua University and the Beijing Natural Science Foundation of China Grant No.~IS25035. SM thanks String Theory group of HRI for useful discussions.

\appendix

\section{Details of the Carroll-Weyl Faddeev-Popov determinant}
\label{app:fp-details}

This appendix gives the algebra behind section \ref{sec:fp-bcs}.  The combined infinitesimal
diffeomorphism and Carroll-Weyl transformations are
\begin{align}
  \delta_\eta V^a
  &=
  -V^b\partial_b\epsilon^a
  +\epsilon^b\partial_bV^a
  +\frac12V^a\partial_b\epsilon^b
  -\lambda V^a,
  \label{app-deltaV}
  \\
  \delta_\eta W_a
  &=
  \epsilon^b\partial_bW_a+W_b\partial_a\epsilon^b-\partial_a\lambda .
  \label{app-deltaW}
\end{align}
The gauge slice is
\begin{equation}
  V^0=1,\qquad V^1=0,\qquad \cW=V^aW_a=W_0=0,
\end{equation}
with \(W_1\) left arbitrary because it does not enter the action through \(V^aW_a\).

For \(G^0=V^0-1\), we have
\begin{align}
  \delta G^0
  &=
  \delta V^0
  \notag\\
  &=
  -V^b\partial_b\epsilon^0
  +\epsilon^b\partial_bV^0
  +\frac12V^0(\partial_0\epsilon^0+\partial_1\epsilon^1)
  -\lambda V^0
  \notag\\
  &=
  -\partial_0\epsilon^0
  +\frac12(\partial_0\epsilon^0+\partial_1\epsilon^1)
  -\lambda
  \notag\\
  &=
  -\frac12\partial_0\epsilon^0+\frac12\partial_1\epsilon^1-\lambda .
  \label{app-dG0}
\end{align}
For \(G^1=V^1\),
\begin{align}
  \delta G^1
  &=
  \delta V^1
  \notag\\
  &=
  -V^b\partial_b\epsilon^1
  +\epsilon^b\partial_bV^1
  +\frac12V^1(\partial_0\epsilon^0+\partial_1\epsilon^1)
  -\lambda V^1
  \notag\\
  &=
  -\partial_0\epsilon^1 .
  \label{app-dG1}
\end{align}
For \(G^s=\cW=V^aW_a\), one must vary both \(V^a\) and \(W_a\)
\begin{align}
  \delta G^s
  &=
  \delta(V^aW_a)
  =
  W_a\delta V^a+V^a\delta W_a
  \notag\\
  &=
  W_0\delta V^0+W_1\delta V^1+\delta W_0 .
  \label{app-dGs-start}
\end{align}
On the slice \(W_0=0\), the first term vanishes.  Since \(W_0(\tau,\sigma)=0\) as a gauge condition,
\(\partial_aW_0=0\) on the slice, and \eqref{app-deltaW} gives
\begin{equation}
  \delta W_0
  =
  W_1\partial_0\epsilon^1-\partial_0\lambda .
\end{equation}
Using \(\delta V^1=-\partial_0\epsilon^1\), the \(W_1\)-dependence cancels
\begin{align}
  \delta G^s
  &=
  W_1(-\partial_0\epsilon^1)
  +W_1\partial_0\epsilon^1
  -\partial_0\lambda
  =
  -\partial_0\lambda .
  \label{app-dGs}
\end{align}
Equations \eqref{app-dG0}, \eqref{app-dG1}, and \eqref{app-dGs} give
\begin{equation}
  \begin{pmatrix}
    \delta G^0\\
    \delta G^1\\
    \delta G^s
  \end{pmatrix}
  =
  \begin{pmatrix}
    -\frac12\partial_0 & \frac12\partial_1 & -1\\
    0 & -\partial_0 & 0\\
    0 & 0 & -\partial_0
  \end{pmatrix}
  \begin{pmatrix}
    \epsilon^0\\
    \epsilon^1\\
    \lambda
  \end{pmatrix}.
  \label{app-M-again}
\end{equation}
This is the operator \(\cM\) in \eqref{Mmatrix}. This matrix is the local Faddeev--Popov operator for the complete Carroll-Weyl gauge fixing.  The entry $-\frac12\partial_0$ and $\frac12\partial_1$ in the first row come from the diffeomorphism variation of $V^0$, while the algebraic entry $-1$ comes from the Carroll-Weyl scaling of $V^0$.  The second row is the usual temporal-gauge condition on $V^1$.  The third row follows from varying the invariant combination $\cW=V^aW_a$; the apparent dependence on the unfixed component $W_1$ cancels between $\delta V^1$ and $\delta W_0$, leaving only $-\partial_0\lambda$.  This cancellation is a useful consistency check, because $W_1$ does not enter the gauge-fixed action through $V^aW_a$.  Therefore the operator $\cM$ captures exactly the three independent gauge directions that must be divided out in the Carroll-Weyl covariant path integral.

The Grassmann representation of the determinant is the continuum version of the finite-dimensional
identity
\begin{equation}
  \det M
  =
  \int \prod_A db_A\,dc^A\,
  \exp\!\left(-b_AM^A{}_Bc^B\right).
\end{equation}
With \(c^A=(c^0,c^1,s)\) and \(b_A=(2b^0,2b^1,b^s)\),
\begin{align}
  -b_A\cM^A{}_Bc^B
  &=
  -2b^0\left(-\frac12\partial_0c^0+\frac12\partial_1c^1-s\right)
  -2b^1(-\partial_0c^1)
  -b^s(-\partial_0s)
  \notag\\
  &=
  b^0\partial_0c^0-b^0\partial_1c^1+2b^0s
  +2b^1\partial_0c^1+b^s\partial_0s .
  \label{app-preparts}
\end{align}
On the cylinder \(\Sigma_{\mathrm{cyl}}\), all total derivatives vanish under the boundary conditions
stated below \eqref{worldsheet-cylinder}.  For two Grassmann-odd fields \(u,v\),
\begin{equation}
  0
  =
  \int_{\Sigma_{\mathrm{cyl}}}\dd^2\sigma\,\partial_a(uv)
  =
  \int_{\Sigma_{\mathrm{cyl}}}\dd^2\sigma\,
  \left((\partial_au)v+u\partial_av\right)
  =
  \int_{\Sigma_{\mathrm{cyl}}}\dd^2\sigma\,
  \left(-v\partial_au+u\partial_av\right),
\end{equation}
so
\begin{equation}
  \int_{\Sigma_{\mathrm{cyl}}}\dd^2\sigma\,u\partial_av
  =
  \int_{\Sigma_{\mathrm{cyl}}}\dd^2\sigma\,v\partial_au .
\end{equation}
Applying this identity to \eqref{app-preparts}, and using \(b^0s=-sb^0\), gives precisely the
integrated form \eqref{bcs-action}.

\section{Component derivation of the \texorpdfstring{\(bcs\)}{bcs} ghost action}
\label{app:ghost-action-details}

This appendix isolates the derivation of the ghost action from the determinant.  The Faddeev-Popov
operator in \eqref{Mmatrix} should be understood as a differential operator acting on ghosts to its
right
\begin{equation}
  (\cM c)^A(\sigma)
  =
  \int\dd^2\sigma'\,
  \cM^A{}_B(\sigma,\sigma')c^B(\sigma'),
  \qquad
  c^B=(c^0,c^1,s).
\end{equation}
Equivalently, suppressing the delta functions in the kernel,
\begin{align}
  (\cM c)^0
  &=
  -\frac12\partial_0c^0+\frac12\partial_1c^1-s,
  \label{app-Mc0}
  \\
  (\cM c)^1
  &=
  -\partial_0c^1,
  \label{app-Mc1}
  \\
  (\cM c)^s
  &=
  -\partial_0s.
  \label{app-Mcs}
\end{align}
The phase convention used in the Lorentzian path integral is
\begin{equation}
  \det\cM
  =
  \int Db_A\,Dc^A\,
  \exp\left\{
  -i\int\dd^2\sigma\,b_A(\cM c)^A
  \right\}
  =
  \int Db_A\,Dc^A\,\exp\{iS_{\mathrm{gh}}\},
  \label{app-det-phase}
\end{equation}
so
\begin{equation}
  S_{\mathrm{gh}}
  =
  -\int\dd^2\sigma\,b_A(\cM c)^A .
  \label{app-Sgh-raw}
\end{equation}
The antighost normalization
\begin{equation}
  b_A=(2b^0,2b^1,b^s)
  \label{app-antighost-normalization}
\end{equation}
is chosen so that the first two rows reproduce the standard BMS \(bc\) normalization when the
Carroll-Weyl ghost is omitted.  Substituting \eqref{app-Mc0}--\eqref{app-Mcs} and
\eqref{app-antighost-normalization} into \eqref{app-Sgh-raw} gives
\begin{align}
  S_{\mathrm{gh}}
  &=
  -\int\dd^2\sigma\,
  \left[
  2b^0\left(-\frac12\partial_0c^0+\frac12\partial_1c^1-s\right)
  +2b^1(-\partial_0c^1)
  +b^s(-\partial_0s)
  \right]
  \notag\\
  &=
  \int\dd^2\sigma\,
  \left[
  b^0\partial_0c^0
  -b^0\partial_1c^1
  +2b^0s
  +2b^1\partial_0c^1
  +b^s\partial_0s
  \right].
  \label{app-Sgh-before-parts}
\end{align}
The integration-by-parts convention is fixed by the cylinder \(\Sigma_{\mathrm{cyl}}\) in
\eqref{worldsheet-cylinder}.  For Grassmann-odd fields \(u,v\),
\begin{equation}
  \int_{\Sigma_{\mathrm{cyl}}}\dd^2\sigma\,u\,\partial_av
  =
  \int_{\Sigma_{\mathrm{cyl}}}\dd^2\sigma\,v\,\partial_au,
  \label{app-odd-ibp}
\end{equation}
because the total derivative of \(uv\) integrates to zero and \((\partial_au)v=-v\partial_au\).
Using \eqref{app-odd-ibp},
\begin{align}
  \int b^0\partial_0c^0
  &=
  \int c^0\partial_0b^0,
  &
  -\int b^0\partial_1c^1
  &=
  -\int c^1\partial_1b^0,
  \notag\\
  2\int b^1\partial_0c^1
  &=
  2\int c^1\partial_0b^1,
  &
  \int b^s\partial_0s
  &=
  \int s\partial_0b^s.
  \label{app-termwise-ibp}
\end{align}
The non-derivative term is reordered by fermionic anticommutation,
\begin{equation}
  2b^0s=-2sb^0.
  \label{app-algebraic-reorder}
\end{equation}
Combining \eqref{app-Sgh-before-parts}, \eqref{app-termwise-ibp}, and
\eqref{app-algebraic-reorder}, and restoring the common factor \(i/(2\pi)\) in the Lorentzian action,
one obtains
\begin{equation}
  S_{\mathrm{gh}}
  =
  \frac{1}{2\pi}\int\dd^2\sigma\,
  i\left(
  c^0\partial_0b^0
  -c^1\partial_1b^0
  +2c^1\partial_0b^1
  +s\partial_0b^s
  -2sb^0
  \right).
  \label{app-Sgh-final}
\end{equation}
Adding the gauge-fixed matter action \(\frac{1}{2\pi}\int\dd^2\sigma\,\dot X^2\) gives
\eqref{bcs-action}.

There is a useful technical point about \(\cM_{\CW}\).  Since the matrix is triangular, its formal
determinant factorizes as in \eqref{triangular-det}.  However, replacing \(\cM_{\CW}\) by only its
diagonal part would not be the same local gauge-fixed theory.  Diagonalizing the first row would require
field redefinitions involving inverse time derivatives, schematically \(\partial_0^{-1}s\), and such
redefinitions are sensitive to zero modes on the cylinder.  More importantly, the off-diagonal
\(-1\) is the BRST variation of the gauge condition \(G^0=V^0-1\) under a Carroll-Weyl transformation
\begin{equation}
  \delta_\lambda G^0=-\lambda .
\end{equation}
The local \(s\,b^0\) term is therefore the representative of the full gauge complex before any
nonlocal diagonalization.  This is the representative appropriate for constructing the BRST charge,
ghost number, vertex operators, and zero-mode measure.

\section{Details of the \texorpdfstring{\(bcs\)}{bcs} equations and modes}
\label{app:modes-details}

The ghost part of the action is
\begin{equation}
  S_{bcs}
  =
  \frac{i}{2\pi}\int\dd^2\sigma\,
  \left(
  c^0\partial_0b^0
  -c^1\partial_1b^0
  +2c^1\partial_0b^1
  +s\partial_0b^s
  -2sb^0
  \right).
\end{equation}
Varying \(c^0,c^1,s\) gives
\begin{align}
  \delta_{c^0}S_{bcs}
  &=
  \frac{i}{2\pi}\int\dd^2\sigma\,\delta c^0\,\partial_0b^0,
  &
  &\Rightarrow\quad
  \partial_0b^0=0,
  \\
  \delta_{c^1}S_{bcs}
  &=
  \frac{i}{2\pi}\int\dd^2\sigma\,\delta c^1
  \left(-\partial_1b^0+2\partial_0b^1\right),
  &
  &\Rightarrow\quad
  \frac12\partial_1b^0-\partial_0b^1=0,
  \\
  \delta_sS_{bcs}
  &=
  \frac{i}{2\pi}\int\dd^2\sigma\,\delta s
  \left(\partial_0b^s-2b^0\right),
  &
  &\Rightarrow\quad
  \partial_0b^s-2b^0=0 .
\end{align}
Varying \(b^0,b^1,b^s\) gives, after Grassmann integration by parts,
\begin{align}
  \delta_{b^0}S_{bcs}
  &=
  \frac{i}{2\pi}\int\dd^2\sigma\,\delta b^0
  \left(\partial_0c^0-\partial_1c^1+2s\right),
  &
  &\Rightarrow\quad
  \partial_0c^0-\partial_1c^1+2s=0,
  \\
  \delta_{b^1}S_{bcs}
  &=
  \frac{i}{2\pi}\int\dd^2\sigma\,2\delta b^1\,\partial_0c^1,
  &
  &\Rightarrow\quad
  \partial_0c^1=0,
  \\
  \delta_{b^s}S_{bcs}
  &=
  \frac{i}{2\pi}\int\dd^2\sigma\,\delta b^s\,\partial_0s,
  &
  &\Rightarrow\quad
  \partial_0s=0 .
\end{align}

The mode expansion follows by integrating these first-order equations.  Periodicity on \(S^1\) implies
Fourier expansions in \(e^{-in\sigma}\).  Thus
\begin{equation}
  c^1(\tau,\sigma)=\sum_n c_ne^{-in\sigma},
  \qquad
  s(\tau,\sigma)=\sum_n s_ne^{-in\sigma}.
\end{equation}
Then
\begin{equation}
  \partial_0c^0
  =
  \partial_1c^1-2s
  =
  \sum_n(-in c_n-2s_n)e^{-in\sigma},
\end{equation}
so
\begin{equation}
  c^0(\tau,\sigma)
  =
  \sum_n(\tilde c_n-in\tau c_n-2\tau s_n)e^{-in\sigma}.
\end{equation}
Similarly
\begin{equation}
  b^0(\tau,\sigma)=\sum_n\tilde b_ne^{-in\sigma},
\end{equation}
and
\begin{equation}
  \partial_0b^1
  =
  \frac12\partial_1b^0
  =
  -\frac{i}{2}\sum_n n\tilde b_ne^{-in\sigma},
\end{equation}
which integrates to
\begin{equation}
  b^1(\tau,\sigma)
  =
  \frac12\sum_n(b_n-in\tau\tilde b_n)e^{-in\sigma}.
\end{equation}
Finally,
\begin{equation}
  \partial_0b^s=2b^0
  =
  2\sum_n\tilde b_ne^{-in\sigma},
\end{equation}
so
\begin{equation}
  b^s(\tau,\sigma)
  =
  \sum_n(r_n+2\tau\tilde b_n)e^{-in\sigma}.
\end{equation}

At \(\tau=0\), the ghost symplectic one-form is
\begin{equation}
  \Theta_{bcs}
  =
  \frac{i}{2\pi}\int_0^{2\pi}\dd\sigma\,
  \left(c^0\delta b^0+2c^1\delta b^1+s\delta b^s\right)
  =
  i\sum_n\left(
  \tilde c_{-n}\delta\tilde b_n
  +c_{-n}\delta b_n
  +s_{-n}\delta r_n
  \right).
\end{equation}
This gives the canonical anticommutators in \eqref{bcs-anticommutators}.

\section{Derivation of the ghost oscillator algebra} \label{app:oscillatoralgebra}

In this appendix we discuss the mode algebra \eqref{bcs-anticommutators} directly from the
ghost action. The ghost sector of the gauge-fixed action is

\begin{equation}
S_{bcs}
=
\frac{i}{2\pi}
\int d^{2}\sigma
\left(
c^{0}\partial_{0}b^{0}
-c^{1}\partial_{1}b^{0}
+2c^{1}\partial_{0}b^{1}
+s\partial_{0}b^{s}
-2sb^{0}
\right).
\label{D.1}
\end{equation}

Only the terms containing time derivatives contribute to the symplectic
structure. The algebraic mixing term $-2sb^{0}$ contains no
$\tau$-derivative and therefore does not affect the equal-time
canonical brackets. The kinetic part of the action is

\begin{equation}
S_{\rm kin}
=
\frac{i}{2\pi}
\int d^{2}\sigma
\left(
c^{0}\partial_{0}b^{0}
+2c^{1}\partial_{0}b^{1}
+s\partial_{0}b^{s}
\right).
\label{D.2}
\end{equation}

The canonical momenta conjugate to the antighost fields are
\begin{align}
\pi_{b^{0}}
&=
\frac{\partial \mathcal L}{\partial(\partial_{0}b^{0})}
=
\frac{i}{2\pi}c^{0},
\label{D.3}
\\[2mm]
\pi_{b^{1}}
&=
\frac{\partial \mathcal L}{\partial(\partial_{0}b^{1})}
=
\frac{i}{\pi}c^{1},
\label{D.4}
\\[2mm]
\pi_{b^{s}}
&=
\frac{\partial \mathcal L}{\partial(\partial_{0}b^{s})}
=
\frac{i}{2\pi}s.
\label{D.5}
\end{align}

The equal-time graded Poisson brackets therefore take the form
\begin{align}
\{b^{0}(\sigma),c^{0}(\sigma')\}
&=
2\pi\,\delta(\sigma-\sigma'),
\label{D.6}
\\
\{b^{1}(\sigma),c^{1}(\sigma')\}
&=
\pi\,\delta(\sigma-\sigma'),
\label{D.7}
\\
\{b^{s}(\sigma),s(\sigma')\}
&=
2\pi\,\delta(\sigma-\sigma').
\label{D.8}
\end{align}
Notice that the factor of $\pi$ instead of $2\pi$ in \eqref{D.7} arises from the factor $2$ multiplying $c^{1}\partial_{0}b^{1}$ in the kinetic term.\\

We now substitute the mode expansions
\begin{align}
c^{1}(\tau,\sigma)
&=
\sum_{n} c_{n}e^{-in\sigma},
\label{D.9}
\\
s(\tau,\sigma)
&=
\sum_{n} s_{n}e^{-in\sigma},
\label{D.10}
\\
c^{0}(\tau,\sigma)
&=
\sum_{n}
(\tilde c_{n}-in\tau c_{n}-2\tau s_{n})
e^{-in\sigma},
\label{D.11}
\\
b^{0}(\tau,\sigma)
&=
\sum_{n}\tilde b_{n}e^{-in\sigma},
\label{D.12}
\\
b^{1}(\tau,\sigma)
&=
\frac12
\sum_{n}
(b_{n}-in\tau\tilde b_{n})
e^{-in\sigma},
\label{D.13}
\\
b^{s}(\tau,\sigma)
&=
\sum_{n}
(r_{n}+2\tau\tilde b_{n})
e^{-in\sigma}.
\label{D.14}
\end{align}

Using
\begin{equation}
\delta(\sigma-\sigma')
=
\frac{1}{2\pi}
\sum_{k}
e^{-ik(\sigma-\sigma')},
\label{D.15}
\end{equation}

we first consider the pair $(b_{n},c_{n})$. Substituting
(\ref{D.9}) and (\ref{D.13}) into (\ref{D.7}) gives

\begin{equation}
\frac12
\sum_{m,n}
\{b_{m},c_{n}\}
e^{-im\sigma}
e^{-in\sigma'}
=
\pi\delta(\sigma-\sigma')
=
\frac12
\sum_{k}
e^{-ik\sigma}e^{ik\sigma'}.
\label{D.16}
\end{equation}

Matching Fourier coefficients yields

\begin{equation}
\{b_{m},c_{n}\}
=
\delta_{m+n,0}.
\label{D.17}
\end{equation}

Next consider the pair
$(\tilde b_{n},\tilde c_{n})$. Using
(\ref{D.6}), (\ref{D.11}) and (\ref{D.12}) we obtain
\begin{equation}
\sum_{m,n}
\{\tilde b_{m},\tilde c_{n}\}
e^{-im\sigma}
e^{-in\sigma'}
=
\sum_{k}
e^{-ik\sigma}e^{ik\sigma'}.
\label{D.18}
\end{equation}

The terms proportional to $c_{n}$ and $s_{n}$ do not contribute,
since they belong to different canonical sectors. Therefore
\begin{equation}
\{\tilde b_{m},\tilde c_{n}\}
=
\delta_{m+n,0}.
\label{D.19}
\end{equation}

Finally, using (\ref{D.8}), (\ref{D.10}) and (\ref{D.14}),

\begin{equation}
\sum_{m,n}
\{r_{m},s_{n}\}
e^{-im\sigma}
e^{-in\sigma'}
=
\sum_{k}
e^{-ik\sigma}e^{ik\sigma'}.
\label{D.20}
\end{equation}

The $\tilde b_{n}$ contribution in (\ref{D.14}) again drops out
because it belongs to a different canonical pair. Hence
\begin{equation}
\{r_{m},s_{n}\}
=
\delta_{m+n,0}.
\label{D.21}
\end{equation}

Combining (\ref{D.17}), (\ref{D.19}) and (\ref{D.21}) gives the
oscillator algebra
\begin{equation}
\{b_{n},c_{m}\}
=
\delta_{m+n,0},
\qquad
\{\tilde b_{n},\tilde c_{m}\}
=
\delta_{m+n,0},
\qquad
\{r_{n},s_{m}\}
=
\delta_{m+n,0}.
\label{D.22}
\end{equation}

This reproduces the mode algebra quoted in \eqref{bcs-anticommutators}. The
$\tau$-dependent terms appearing in $c^{0}$ and $b^{s}$ modify the
solutions of the equations of motion but not the symplectic form,
since they arise from the Hamiltonian evolution generated by the
mixing term $-2sb^{0}$ and do not introduce additional kinetic
couplings.

\section{Semidirect products and the Carroll-Weyl gauge group}
\label{app:semidirect}

In this appendix we explain why the gauge group of the Carroll-Weyl covariant null string is written as
\(
\mathrm{Diff}\ltimes \mathrm{CW},
\)
rather than the direct product
\(
\mathrm{Diff}\times \mathrm{CW}.
\)
The distinction is important because it is reflected directly in the
constraint algebra and in the structure of the Faddeev-Popov
determinant.

\subsection*{Direct and semidirect products}

Suppose that a theory possesses two symmetry groups \(G\) and \(H\).
If the two groups act independently, the full symmetry group is the
direct product
\begin{equation}
G\times H
\label{eq:sdp_direct}
\end{equation}
At the infinitesimal level this means that generators belonging to the
two sectors commute,
\begin{equation}
[T_G,T_H]=0
\label{eq:sdp_commuting}
\end{equation}
A semidirect product arises when one subgroup acts nontrivially on the
other. In this case the full symmetry group is written
\begin{equation}
G\ltimes H
\label{eq:sdp_semidirect}
\end{equation}
The generators no longer commute. Instead,
\begin{equation}
[T_G,T_H]\sim T_H
\label{eq:sdp_noncommuting}
\end{equation}
showing that transformations generated by \(G\) rotate or mix the
generators belonging to \(H\).\\

A familiar example is the Euclidean group
\begin{equation}
ISO(d)=SO(d)\ltimes \mathbb{R}^{d}
\label{eq:sdp_iso}
\end{equation}
The translation generators \(P_i\) transform under rotations \(J_{ij}\),
\begin{equation}
[J_{ij},P_k]
=
\delta_{jk}P_i-\delta_{ik}P_j
\label{eq:sdp_iso_alg}
\end{equation}
Thus rotations act nontrivially on translations, and the symmetry group
is a semidirect product rather than a direct product.

\subsection*{Semidirect products in gauge theories}

The same phenomenon appears in local gauge theories whenever one gauge
transformation acts on the parameter of another gauge transformation.

Consider two local gauge symmetries with parameters \(\xi\) and
\(\lambda\). If under a \(\xi\)-transformation the second parameter
changes according to
\begin{equation}
\delta_\xi \lambda \neq 0
\label{eq:sdp_param_transform}
\end{equation}
then the two gauge sectors are not independent. Acting with a
\(\xi\)-transformation modifies a subsequent \(\lambda\)-transformation.

At the infinitesimal level one finds
\begin{equation}
[\delta_\xi,\delta_\lambda]
=
\delta_{\delta_\xi\lambda}
\label{eq:sdp_gauge_comm}
\end{equation}
which is again a transformation belonging to the \(\lambda\)-sector.
The gauge group therefore acquires the structure of a semidirect
product.

The essential criterion is simple: whenever one gauge transformation
acts nontrivially on the parameter of another gauge transformation, the
full gauge group is generically a semidirect product.

\subsection*{Diffeomorphisms and Weyl symmetry in the Polyakov string}

The Polyakov action is invariant under worldsheet diffeomorphisms and
Weyl transformations,
\begin{equation}
\delta_\xi g_{ab}
=
\nabla_a\xi_b+\nabla_b\xi_a
\label{eq:sdp_diff_metric}
\end{equation}
and
\begin{equation}
\delta_\omega g_{ab}
=
2\omega g_{ab}
\label{eq:sdp_weyl_metric}
\end{equation}

The Weyl parameter \(\omega(\sigma)\) is a local scalar function on the
worldsheet. Under a diffeomorphism generated by \(\xi^a\),
\begin{equation}
\delta_\xi\omega
=
\xi^a\partial_a\omega
\label{eq:sdp_weyl_param}
\end{equation}

Consequently,
\begin{equation}
[\delta_\xi,\delta_\omega]g_{ab}
=
2(\xi^c\partial_c\omega)g_{ab}
\label{eq:sdp_diff_weyl_comm}
\end{equation}
which may be rewritten as
\begin{equation}
[\delta_\xi,\delta_\omega]
=
\delta_{\xi\cdot\partial\omega}
\label{eq:sdp_diff_weyl_closure}
\end{equation}

Thus diffeomorphisms act nontrivially on local Weyl transformations.
Strictly speaking, the local gauge group is therefore
\begin{equation}
\mathrm{Diff}\ltimes \mathrm{Weyl}
\label{eq:sdp_diff_weyl_group}
\end{equation}

In practice this distinction is rarely emphasized. After conformal
gauge fixing the Weyl factor is removed through the trace part of the
metric variation, and the resulting Faddeev-Popov determinant is
represented solely by the usual \(bc\)-ghost system. Weyl symmetry does
not introduce an additional independent first-class constraint in the
physical phase space.

\subsection*{Diffeomorphisms and Carroll-Weyl symmetry in the null string}

The Carroll-Weyl covariant null string possesses local
diffeomorphisms together with local Carroll-Weyl transformations.
The latter act on the embedding coordinates as
\begin{equation}
\delta_\chi X^\mu
=
\chi X^\mu
\label{eq:sdp_cw_X}
\end{equation}

The Carroll-Weyl parameter \(\chi(\tau,\sigma)\) is again a local
worldsheet scalar. Under diffeomorphisms,
\begin{equation}
\delta_\xi\chi
=
\xi^a\partial_a\chi
\label{eq:sdp_cw_param}
\end{equation}

Applying the commutator to the embedding coordinates gives
\begin{equation}
[\delta_\xi,\delta_\chi]X^\mu
=
(\xi^a\partial_a\chi)X^\mu
\label{eq:sdp_diff_cw_comm}
\end{equation}

Equivalently,
\begin{equation}
[\delta_\xi,\delta_\chi]
=
\delta_{\xi\cdot\partial\chi}
\label{eq:sdp_diff_cw_closure}
\end{equation}

Hence the Carroll-Weyl subgroup is acted upon nontrivially by
diffeomorphisms, and the gauge group takes the form
\begin{equation}
\mathrm{Diff}\ltimes \mathrm{CW}
\label{eq:sdp_diff_cw_group}
\end{equation}

Unlike the Polyakov string, however, the Carroll-Weyl symmetry is
generated by an independent first-class constraint,
\begin{equation}
C_3=P\cdot X
\label{eq:sdp_cw_constraint}
\end{equation}
The Carroll-Weyl sector is therefore a genuine part of the physical
gauge algebra rather than merely a redundancy of auxiliary worldsheet
variables.

\subsection*{Constraint algebra interpretation}

The semidirect-product structure is reflected directly in the
constraint algebra. The mode generators satisfy
\begin{equation}
\{L_m,S_n\}
=
in\,S_{m+n}
\label{eq:sdp_LS}
\end{equation}
where \(L_m\) generate superrotations and \(S_n\) generate
Carroll-Weyl transformations.

This relation has the same mathematical form as the Euclidean-group
commutator
\begin{equation}
[J,P]\sim P
\label{eq:sdp_euclidean_structure}
\end{equation}
The \(S_n\) generators transform nontrivially under the action of
\(L_m\) and therefore do not form an independent commuting sector.

Furthermore,
\begin{equation}
\{M_m,S_n\}
=
-2M_{m+n}
\label{eq:sdp_MS}
\end{equation}
showing that Carroll-Weyl transformations act nontrivially on the
supertranslation generators as well.

The constraint algebra therefore realizes the semidirect-product
structure directly at the Hamiltonian level.

\subsection*{Implications for the Faddeev-Popov construction}

The importance of the semidirect-product structure is that the
Carroll-Weyl symmetry cannot be treated as a spectator gauge factor.
Since \(C_3=P\cdot X\) is an independent first-class constraint, the
path integral must divide by the full gauge orbit generated by
\begin{equation}
\mathrm{Diff}\ltimes \mathrm{CW}
\label{eq:sdp_final_group}
\end{equation}

The resulting Faddeev-Popov operator therefore contains three gauge
directions rather than two. This enlargement introduces the additional
ghost-antighost pair \((s,b_s)\) and leads to the \(bcs\)-ghost system
derived in Section~\ref{sec:fp-bcs}.

From this viewpoint, the off-diagonal entries of the Faddeev-Popov
operator and the mixing term \(sb_0\) are the path-integral
manifestation of the semidirect-product structure of the gauge group.

\section{Grassmann proof of the fermionic determinant identity}
\label{app:grassmann-det}

In this appendix we prove the finite-dimensional Grassmann identity
\begin{equation}
  \det M
  =
  \int \prod_A db_A\,dc^A\,
  \exp\!\left[-b_AM^A{}_Bc^B\right],
  \label{finite-grassmann-det-proof}
\end{equation}
up to an overall convention-dependent normalization of the Berezin measure.
Here $M^A{}_B$ is an $N\times N$ matrix, and $b_A,c^A$ are independent
Grassmann variables,
\begin{equation}
  b_A b_B=-b_B b_A,\qquad
  c^A c^B=-c^B c^A,\qquad
  b_A c^B=-c^B b_A .
\end{equation}
We choose the orientation of the Berezin measure such that
\begin{equation}
  \int \prod_A db_A\,dc^A\,
  c^1 b_1 c^2 b_2\cdots c^N b_N
  =1 .
  \label{measure-orientation}
\end{equation}
A different ordering changes the result only by an overall sign independent
of $M$.

Using the Grassmann anticommutation relation, the exponent can be written as
\begin{equation}
  -b_AM^A{}_Bc^B
  =
  c^B M^A{}_B b_A .
\end{equation}
Since the Grassmann variables are nilpotent, the exponential terminates
\begin{equation}
  \exp\!\left[-b_AM^A{}_Bc^B\right]
  =
  \sum_{k=0}^{N}
  \frac{1}{k!}
  \left(c^B M^A{}_B b_A\right)^k .
\end{equation}
The Berezin integral is nonzero only for the term containing all $N$
variables $c^A$ and all $N$ variables $b_A$. Hence only the $k=N$ term
contributes
\begin{align}
  \int \prod_A db_A\,dc^A\,
  \exp\!\left[-b_AM^A{}_Bc^B\right]
  &=
  \frac{1}{N!}
  M^{A_1}{}_{B_1}\cdots M^{A_N}{}_{B_N}
  \notag\\
  &\qquad\times
  \int \prod_A db_A\,dc^A\,
  c^{B_1}b_{A_1}\cdots c^{B_N}b_{A_N}.
  \label{top-term}
\end{align}
By definition of the Berezin integral, the last integral is completely
antisymmetric separately in the upper indices $B_i$ and in the lower indices
$A_i$. With the convention \eqref{measure-orientation}, one has
\begin{equation}
  \int \prod_A db_A\,dc^A\,
  c^{B_1}b_{A_1}\cdots c^{B_N}b_{A_N}
  =
  \epsilon^{B_1\cdots B_N}\epsilon_{A_1\cdots A_N}.
  \label{epsilon-integral}
\end{equation}
Substituting this into \eqref{top-term}, we obtain
\begin{equation}
  \int \prod_A db_A\,dc^A\,
  \exp\!\left[-b_AM^A{}_Bc^B\right]
  =
  \frac{1}{N!}
  \epsilon_{A_1\cdots A_N}
  \epsilon^{B_1\cdots B_N}
  M^{A_1}{}_{B_1}\cdots M^{A_N}{}_{B_N}.
\end{equation}
The right-hand side is precisely the standard epsilon-symbol expression for
the determinant
\begin{equation}
  \det M
  =
  \frac{1}{N!}
  \epsilon_{A_1\cdots A_N}
  \epsilon^{B_1\cdots B_N}
  M^{A_1}{}_{B_1}\cdots M^{A_N}{}_{B_N}.
\end{equation}
Therefore,
\begin{equation}
  \det M
  =
  \int \prod_A db_A\,dc^A\,
  \exp\!\left[-b_AM^A{}_Bc^B\right],
\end{equation}
with the Berezin-measure orientation chosen as in
\eqref{measure-orientation}.

Equivalently, for a different ordering of the Berezin measure, one obtains
\begin{equation}
  \int \prod_A db_A\,dc^A\,
  \exp\!\left[-b_AM^A{}_Bc^B\right]
  =
  \pm \det M .
\end{equation}
The sign is independent of $M$ and can be absorbed into the normalization of
the path-integral measure. This is the finite-dimensional origin of the
Faddeev--Popov representation of a determinant by fermionic ghosts.
\bibliographystyle{JHEP}
\bibliography{ref}

@book{TextbookStringTheory,
  author    = {Polchinski, Joseph},
  title     = {String Theory. Vol. 1: An Introduction to the Bosonic String},
  publisher = {Cambridge University Press},
  year      = {1998}
}

@article{Isberg:1993av,
  author        = {Isberg, J. and Lindstrom, U. and Sundborg, B. and Theodoridis, G.},
  title         = {Classical and quantized tensionless strings},
  journal       = {Nucl. Phys. B},
  volume        = {411},
  pages         = {122--156},
  year          = {1994},
  doi           = {10.1016/0550-3213(94)90056-6},
  eprint        = {hep-th/9307108},
  archivePrefix = {arXiv}
}

@article{Chen:2023esw,
  author        = {Chen, Bin and Hu, Zezhou and Yu, Zhe-fei and Zheng, Yu-fan},
  title         = {Path-integral quantization of tensionless (super) string},
  journal       = {JHEP},
  volume        = {08},
  pages         = {133},
  year          = {2023},
  doi           = {10.1007/JHEP08(2023)133},
  eprint        = {2302.05975},
  archivePrefix = {arXiv},
  primaryClass  = {hep-th}
}

@article{Gross:1987kza,
  author  = {Gross, David J. and Mende, Paul F.},
  title   = {The High-Energy Behavior of String Scattering Amplitudes},
  journal = {Phys. Lett. B},
  volume  = {197},
  pages   = {129--134},
  year    = {1987},
  doi     = {10.1016/0370-2693(87)90355-8}
}

@article{Gross:1987ar,
  author  = {Gross, David J. and Mende, Paul F.},
  title   = {String Theory Beyond the Planck Scale},
  journal = {Nucl. Phys. B},
  volume  = {303},
  pages   = {407--454},
  year    = {1988},
  doi     = {10.1016/0550-3213(88)90390-2}
}

@article{Gross:1988ue,
  author  = {Gross, David J.},
  title   = {High-Energy Symmetries of String Theory},
  journal = {Phys. Rev. Lett.},
  volume  = {60},
  pages   = {1229},
  year    = {1988},
  doi     = {10.1103/PhysRevLett.60.1229}
}

@article{Sundborg:2000wp,
  author        = {Sundborg, Bo},
  title         = {Stringy gravity, interacting tensionless strings and massless higher spins},
  journal       = {Nucl. Phys. B Proc. Suppl.},
  volume        = {102},
  pages         = {113--119},
  year          = {2001},
  doi           = {10.1016/S0920-5632(01)01546-4},
  eprint        = {hep-th/0103247},
  archivePrefix = {arXiv}
}

@article{Vasiliev:2003ev,
  author        = {Vasiliev, Mikhail A.},
  title         = {Higher spin gauge theories in various dimensions},
  journal       = {PoS},
  volume        = {JHW2003},
  pages         = {003},
  year          = {2003},
  eprint        = {hep-th/0401177},
  archivePrefix = {arXiv}
}

@article{Schild:1976vq,
  author  = {Schild, A.},
  title   = {Classical Null Strings},
  journal = {Phys. Rev. D},
  volume  = {16},
  pages   = {1722},
  year    = {1977},
  doi     = {10.1103/PhysRevD.16.1722}
}

@article{Bagchi:2013bga,
  author        = {Bagchi, Arjun},
  title         = {Tensionless Strings and Galilean Conformal Algebra},
  journal       = {JHEP},
  volume        = {05},
  pages         = {141},
  year          = {2013},
  doi           = {10.1007/JHEP05(2013)141},
  eprint        = {1303.0291},
  archivePrefix = {arXiv},
  primaryClass  = {hep-th}
}

@article{Casali:2016atr,
  author        = {Casali, Eduardo and Tourkine, Piotr},
  title         = {On the null origin of the ambitwistor string},
  journal       = {JHEP},
  volume        = {11},
  pages         = {036},
  year          = {2016},
  doi           = {10.1007/JHEP11(2016)036},
  eprint        = {1606.05636},
  archivePrefix = {arXiv},
  primaryClass  = {hep-th}
}

@article{Casali:2017zkz,
  author        = {Casali, Eduardo and Herfray, Yannick and Tourkine, Piotr},
  title         = {The complex null string, Galilean conformal algebra and scattering equations},
  journal       = {JHEP},
  volume        = {10},
  pages         = {164},
  year          = {2017},
  doi           = {10.1007/JHEP10(2017)164},
  eprint        = {1707.09900},
  archivePrefix = {arXiv},
  primaryClass  = {hep-th}
}

@article{Mason:2013sva,
  author        = {Mason, Lionel and Skinner, David},
  title         = {Ambitwistor strings and the scattering equations},
  journal       = {JHEP},
  volume        = {07},
  pages         = {048},
  year          = {2014},
  doi           = {10.1007/JHEP07(2014)048},
  eprint        = {1311.2564},
  archivePrefix = {arXiv},
  primaryClass  = {hep-th}
}

@article{SheikhJabbari:2026Consistency,
  author        = {Sheikh-Jabbari, M. M. and Yavartanoo, Hossein},
  title         = {On the Consistency of Null Strings Literature: The Tale of an Overlooked Symmetry},
  year          = {2026},
  eprint        = {2605.12414},
  archivePrefix = {arXiv},
  primaryClass  = {hep-th}
}

@article{SheikhJabbari:2026I,
  author        = {Sheikh-Jabbari, M. M. and Yavartanoo, H.},
  title         = {Null Strings Gauged and Reloaded, I: Null Strings Have Carroll-Weyl Gauge Symmetry},
  year          = {2026},
  eprint        = {2605.25817},
  archivePrefix = {arXiv},
  primaryClass  = {hep-th}
}

@article{SheikhJabbari:2026II,
  author        = {Sheikh-Jabbari, M. M. and Yavartanoo, H.},
  title         = {Null Strings Gauged and Reloaded, II: Consistent Classical Treatment of the Null Strings},
  year          = {2026},
  eprint        = {2605.26822},
  archivePrefix = {arXiv},
  primaryClass  = {hep-th}
}

@article{DuaryMaji:2026ToAppear,
  author        = {Duary, Sarthak and Maji, Sourav},
  title         = {Quantum extended BMS algebra and BRST cohomology of the Carroll--Weyl bcs ghost system},
  year          = {2026},
  eprint        = {to appear soon},
  archivePrefix = {arXiv},
  primaryClass  = {hep-th}
}

@article{Isberg:1992ia,
    author = "Isberg, J. and Lindstrom, U. and Sundborg, B.",
    title = "{Space-time symmetries of quantized tensionless strings}",
    eprint = "hep-th/9207005",
    archivePrefix = "arXiv",
    reportNumber = "USITP-92-4",
    doi = "10.1016/0370-2693(92)90890-G",
    journal = "Phys. Lett. B",
    volume = "293",
    pages = "321--326",
    year = "1992"
}

@article{Gustafsson:1994kr,
    author = "Gustafsson, H. and Lindstrom, U. and Saltsidis, P. and Sundborg, B. and van Unge, R.",
    title = "{Hamiltonian BRST quantization of the conformal string}",
    eprint = "hep-th/9410143",
    archivePrefix = "arXiv",
    reportNumber = "USITP-94-08",
    doi = "10.1016/0550-3213(95)00051-S",
    journal = "Nucl. Phys. B",
    volume = "440",
    pages = "495--520",
    year = "1995"
}

@article{Chen:2026tensionlessbranes,
    author = "Chen, Bin and Hu, Zezhou",
    title = "{Symmetries and Critical Dimensions of Tensionless Branes}",
    eprint = "2604.01883",
    archivePrefix = "arXiv",
    primaryClass = "hep-th",
    month = "4",
    year = "2026"
}

@article{Bagchi:2015nca,
    author = "Bagchi, Arjun and Chakrabortty, Shankhadeep and Parekh, Pulastya",
    title = "{Tensionless Strings from Worldsheet Symmetries}",
    eprint = "1507.04361",
    archivePrefix = "arXiv",
    primaryClass = "hep-th",
    reportNumber = "MIT-CTP-4690",
    doi = "10.1007/JHEP01(2016)158",
    journal = "JHEP",
    volume = "01",
    pages = "158",
    year = "2016"
}

@article{Lindstrom:2026quz,
    author = {Lindstr{\"o}m, Ulf},
    title = "{Symmetries of tensionless strings}",
    eprint = "2605.26185",
    archivePrefix = "arXiv",
    primaryClass = "hep-th",
    reportNumber = "Uppsala Institute for Theoretical Physics preprint UUITP-08/26",
    month = "5",
    year = "2026"
}

@article{Karlhede:1986wb, author = {Karlhede, A. and Lindstr\"om, U.}, title = {The Classical Bosonic String in the Zero Tension Limit}, journal = {Class. Quant. Grav.}, volume = {3}, pages = {L73--L75}, year = {1986}, doi = {10.1088/0264-9381/3/4/002} }

\end{document}